# Surface conduction and reduced electrical resistivity in ultrathin noncrystalline NbP semimetal

Asir Intisar Khan[1], Akash Ramdas[2,†], Emily Lindgren[2,3†], Hyun-Mi Kim[4,†], Byoungjun Won[5], Xiangjin Wu[1], Krishna Saraswat[1], Ching-Tzu Chen[6], Yuri Suzuki,[3,7], Felipe H. da Jornada[2,8], Il-Kwon Oh[5,*], Eric Pop[1,2,7,9,*]

[1]Department of Electrical Engineering, Stanford University, Stanford, CA 94305, USA

[2]Department of Materials Science and Engineering, Stanford University, Stanford, CA 94305, USA

[3]Geballe Laboratory for Advanced Materials, Stanford University, Stanford, CA 94305, USA

[4]Korea Electronics Technology Institute, Seongnam-si 13509, Republic of Korea

[5]Department of Intelligence Semiconductor Engineering, Ajou University, Suwon 16499, Republic of Korea

[6]IBM T.J. Watson Research Center, Yorktown Heights, NY 10598, USA

[7]Department of Applied Physics, Stanford University, Stanford, CA 94305, USA

[8]Stanford Institute for Materials & Energy Sciences, SLAC Natl. Accelerator Lab., Menlo Park, CA 94025, USA

[9]Precourt Institute for Energy, Stanford University, Stanford, CA 94305, USA

[†]These authors contributed equally to this work.

[*]E-mail: epop@stanford.edu, ikoh@ajou.ac.kr

**The electrical resistivity of conventional metals, such as copper, is known to increase in thin films due to electron-surface scattering, limiting the performance of metals in nanoscale electronics. Here, we find an unusual reduction of resistivity with decreasing film thickness in niobium phosphide (NbP) semimetal deposited at relatively low temperatures of 400 °C. In films thinner than 5 nm, the room temperature resistivity (~34 microohm-centimeters for 1.5-nm-thick NbP) was up to six times lower than the bulk NbP resistivity, and lower than conventional metals at similar thickness (typically ~100 microohm-centimeters). Remarkably, the NbP films are not crystalline, but display local nanocrystalline, short-range order within an amorphous matrix. Our analysis suggests that the lower effective resistivity is due to conduction via surface channels, together with high surface carrier density and sufficiently good mobility as the film thickness is reduced. These results and the fundamental insights obtained here could enable ultrathin, low-resistivity wires for nanoelectronics, beyond the limitations of conventional metals.**

Ultrathin conductors with low electrical resistivity are needed for hyper-scaled nanoelectronics (*1*): from metal interconnects for dense logic and memory (*2*, *3*) to neuromorphic (*4*) and spintronic devices (*5*, *6*). Low resistivity allows lower voltage drops and lower signal delays, reducing power dissipation at the system level (*7*). However, the resistivity of conventional metals increases in films thinner than the electron mean free path (few tens of nanometers at room temperature) due to electron-surface scattering (*8*). For example, the room temperature resistivity of sub-5 nm thin Cu or Ru films is up to an order of magnitude larger than in bulk films (> 100 nm) (*8*–*10*). High electrical resistivity of ultrathin metals can be a key contributor to energy consumption in dense logic and memory (*11*, *12*) and ultimately can limit the performance of future data-driven applications (*4*).

In this context, topological Weyl semimetals NbP, NbAs, TaP or TaAs (*13–18*) are promising because they could carry current within surface states that are topologically protected from disorder scattering (*19*). Multifold fermion semimetals CoSi, RhSi, AlPt or GaPd have also been theoretically predicted (*20*, *21*) to benefit from surface conduction with suppressed scattering (*20*). In other words, as the thickness of such semimetals is reduced, the surface contribution to conduction (*22*) could lead to a decreased effective resistivity (*12*, *20*, *23*), whereas in conventional metals with nanoscale thickness the electrons undergo more surface scattering (*8*, *11*, *24*). For example, recent measurements of high-quality crystalline NbAs displayed over an order of magnitude reduction in the effective resistivity of nanobelts, reaching ~2 μohm·cm for ~250 nm thickness compared to their bulk single-crystal value of ~35 μohm·cm (*23*), at room temperature.

Surface-dominated transport has also been recently reported in amorphous $Bi_2Se_3$ topological insulator films (> 75 nm thick) without long-range order (*25*), and disordered Weyl semimetal $WTe_x$ films (*26*) have shown good charge-to-spin conversion and electrical conductivity, comparable to those of crystalline $WTe_2$ (*27*). Such experimental demonstrations with amorphous topological insulators suggest the possibility of surface-state conduction in Weyl semimetals even in the absence of long-range order. However, it is not known if disordered or nanocrystalline semimetals in ultrathin films (that is, sub-5 nm) maintain surface-dominated transport and could be used to realize low-resistivity materials, beyond the limitations of conventional metals. Such non-crystalline semimetals are much more likely to be compatible with modern semiconductor processing and ultra-dense future electronics, where limited thermal budgets (< 500 °C) pose challenges for depositing single-crystal materials.

In this work, we uncover a reduction of electrical resistivity in non-crystalline NbP semimetal with decreasing thickness down to ~1.5 nm. Importantly, we find lower effective resistivity in sub-5 nm thin NbP films compared to their bulk crystalline counterparts, which we attribute to a proportionally higher conduction through a surface channel in the ultrathin films.

**Film growth and resistivity**

The NbP films were sputtered on sapphire and other substrates at 400 °C, a temperature compatible with back-end-of-line (BEOL) semiconductor fabrication (*31*). As shown in **Fig. 1A**, a seed layer of Nb was used to reduce the lattice mismatch between the substrate and the NbP films (*32*), and to promote local short-range order, i.e. nanocrystallinity. All samples were capped in situ with a ~3- to 4-nm thick silicon nitride layer to limit surface oxidation (see Supplementary **Materials and Methods**: **Materials Deposition, fig. S1,** and **table S1**). We used high-angle annular dark-field (HAADF) scanning transmission electron microscope (STEM) to image the cross-section of the NbP/Nb thin films, revealing local short-range order and nanocrystallinity within an amorphous matrix in the NbP layer, across various thicknesses (~18 nm in **Fig. 1B,C** and **figs. S2 and S3** and ~1.5 to 4.3 nm in **fig. S4**). Energy-dispersive spectroscopy (EDS) and x-ray spectroscopy (XPS) analysis confirmed the stoichiometry and the uniform distribution of Nb and P within our sputtered NbP samples (**fig. S5**). STEM, EDS and XPS characterization methods are detailed in Supplementary **Materials and Methods: Materials Characterization**.

We measured the in-plane electrical resistivity of our NbP/Nb films and control Nb samples using standard Hall and eddy-current-based contactless methods (*33*), with details given in Supplementary **Materials and Methods: Device Fabrication and Electrical Measurement**. The control Nb samples were prepared with the same deposition conditions as the Nb seed layers beneath the NbP samples. **Figure 1D** shows that the measured total room temperature resistivity of NbP/Nb films decreased from ~200 μohm·cm for ~80-nm-thick NbP to ~51 μohm·cm for ~1.5-nm NbP (all on 4 nm Nb). This resistivity plot includes the electrical and thickness contribution of the 4-nm seed Nb layer. However, the resistivity of our control Nb metal films increased dramatically as their thickness was reduced over the same range.

The measured temperature dependence of total resistivity in **fig. S6A** revealed metallic behavior (resistivity proportional to temperature) in NbP films of 18 nm or thinner, here including the 4 nm Nb seed layer (measured separately in **fig. S6B**). In contrast, an ~80 nm NbP film (also on a 4 nm Nb seed) showed resistivity almost independent of temperature, a signature of disorder or impurity-dominated bulk states (*25*, *34*). The reduced effective resistivity of the thinner NbP films suggested that there may be a nonnegligible contribution from surface carriers to the total conductance of these samples (*29*, *35*, *36*), which is explored in more detail in Fig. 3 below.

**Figure 1E** shows that the unconventional resistivity scaling with thickness in our NbP/Nb film was preserved for varying thicknesses of Nb seed layer (here 4 nm and 1.4 nm). This decreasing resistivity with decreasing film thickness was also observed after the conductance of the thin Nb seed layer (**fig. S6C**) was subtracted from that of the NbP/Nb stack (**fig. S6D**), indicating that the NbP film was responsible for the observed trend in **Fig. 1E**. For comparison, we also prepared Cu/Nb films with similar thickness, and **fig. S7** shows their resistivity increased as their thickness is reduced, both before and after subtracting the conductance contribution of the 4 nm Nb. In other words, the Nb layer did not influence the contrasting resistivity trend observed for NbP versus Cu. **Figure 1E** also revealed that the resistivity of NbP on 1.4 nm Nb seed was higher than for NbP on 4 nm Nb seed, which we attributed to relatively lower strain in NbP with the thicker Nb seed, as discussed further below. The room temperature resistivity of our sub-3 nm thin NbP films on 4 nm Nb seed is < 45 μohm·cm, lower than the crystalline, bulk NbP resistivity of ~60 to 70 μohm·cm (*14*, *32*). The thinnest, 1.5 nm NbP film reached 34 μohm·cm at room temperature (red circles), after subtracting the Nb seed contribution.

**Figure 1F** displays the scaling of room temperature resistivity versus thickness in our nanocrystalline NbP semimetal, revealing a trend unlike traditional metals such as Cu, Nb, Ta, and achieving one of the lowest resistivities at sub-5 nm thickness. We also quantify the total sheet resistance, $R_\square$, versus. thickness of various films in **fig. S8**, including their seed or barrier layers, if any. As total thickness decreased from ~20 nm down to ~5 nm, $R_\square$ of conventional metals increased by 10 to 100 times, but $R_\square$ of topological semimetals increases by less than a factor of 2. Previously, resistivity smaller than the bulk resistivity was detected in NbAs nanobelts (*23*), topological insulators like $Bi_2Se_3$ (*29*) and multifold fermion semimetal CoSi nanowires (*37*), although such films displayed greater crystallinity, greater thicknesses, and were deposited at higher temperature (typically > 600 °C). Multilayer graphene can also reach low resistivity in nanometer-thin films, but only with substantial doping (*38*, *39*) and with high-temperature growth and processing (*40*). In contrast, the low deposition temperature (400 °C) of our nanocrystalline NbP films was compatible with industrial back-end-of-line processes, a key advantage for integration into state-of-the-art nanoelectronics (*31*, *41*).

We also measured low resistivity and a similar resistivity scaling trend in ultrathin NbP films on different substrates, such as MgO and $SiO_2$/Si (**fig. S9A**), as well as with different capping layers including silica and alumina. In terms of stability, uncapped NbP thin films (~2.6 nm) on 4-nm Nb measured in air show < 10% change of resistivity after 4 days versus ~90% change in 4-nm Nb metal films), indicating a lower surface oxidation of NbP (**fig. S9B**). This stability is also promising for interconnect applications.

**Structural studies**

The resistivity of the sub-20-nm NbP thin films on 4-nm Nb seeds was notably lower than that of NbP on the 1.4-nm Nb seed. To understand this difference, we imaged ~2.6-nm thin NbP films on Nb seed layers with 4- and 1.4-nm thicknesses (**Fig. 2, A** and **B**, respectively) using atomic-resolution HAADF-STEM. Zoomed-in STEM images and corresponding diffraction patterns show the presence of similar nanocrystallinity within the amorphous matrices of NbP on both Nb seed layers (also see **fig. S2** for 18 nm NbP films). Both 4-nm and 1.4-nm Nb seed layers were crystalline (**fig. S3**). NbP films were predominantly amorphous with several nanometer-sized crystalline regions, regardless of the Nb seed-layer thickness. Thus, the observed NbP resistivity scaling with thickness (**Fig. 1E**) for varying Nb seed layers would not likely be affected by the microstructure of the NbP films. The average lattice constant of our ~2.6-nm thin NbP film on 4-nm Nb seed layer (**Fig. 2C**) was ~3.34 Å (~3.33 Å for ~18-nm NbP film, **fig. S10A**), a near that of single-crystal NbP (*42*). However, **Fig. 2D** and **fig. S10B** revealed that the NbP film was strained, with higher average lattice constant (~3.41 Å for ~2.6-nm NbP, and ~3.5 Å for ~18-nm NbP) on the 1.4-nm Nb seed layer, which could cause the higher resistivity (*43*, *44*) for ultrathin NbP on 1.4 nm Nb seed layer (**Fig. 1E**).

We further found (**Fig. 2E**) that the epitaxial relationship between the Nb seed and the $Al_2O_3$ substrate was Nb (001) || $Al_2O_3$ (102). The Al in $Al_2O_3$ (102) had a rhombus lattice tilted by 6° compared to the square lattice of the Nb (100) plane. As a result, in-plane misfit strain occurred between the Nb seed and the substrate (**see fig. S11**). Increasing the Nb seed layer thickness generated misfit dislocations within the Nb that released this strain energy. We observed strain release in the films with ~4-nm Nb seed in **Fig. 2F**, where the Nb lattice returned to its cubic structure with nominal lattice constant of ~3.332 Å (*42*). For thinner 1.4-nm Nb seed, the misfit dislocations that could release stress were not observed (**Fig. 2G**). This laterally strained the 1.4-nm Nb seed layer with a lattice constant of ~3.53 Å, near that of the $Al_2O_3$ substrate and so the NbP films on the 1.4-nm Nb seed also display lateral strain (**Fig. 2D, fig. S10B**). The strained NbP/Nb interface could also cause charge scattering, further increasing the resistivity of the tensile NbP films (*43*, *44*) on the 1.4 nm Nb seed (**Fig. 1E**).

**Transport measurements**

As our next step, we wish to understand what causes the unusual resistivity scaling trend (versus thickness) in our NbP semimetal films. Previous reports have suggested surface-dominated conduction in topological insulators ($Bi_2Se_3$) and topological semimetals (TaAs, NbAs) in their crystalline (*23*, *29*, *35*) as well as in amorphous or nanocrystalline $Bi_2Se_3$ (*25*, *34*) films, attributed to topologically protected surface states. As the sample thickness was decreased, conduction dominated by such surface states could explain the reduced resistivity of our thinner NbP films compared to their thicker counterparts. To understand this, we performed temperature-dependent transport measurements for a series of NbP thin films with varying thicknesses (~80 nm to ~4.3 nm) on 4-nm Nb seed using standard Hall bar devices (**Fig. 3A** and Supplementary **Materials and Methods)**.

The unconventional trend of decreasing resistivity with decreasing NbP/Nb sample thickness persisted across all temperatures probed down to 5 K (**Fig. 3B**). The three thinner NbP/Nb films (4.3-, 9-, and 18-nm NbP, each on 4-nm Nb) showed decreasing resistivity with decreasing temperature (metallic behavior). In contrast, the thick NbP/Nb film (~80-nm NbP on 4-nm Nb seed) displayed a resistivity almost independent of temperature, a signature of disorder- or impurity-dominated bulk states (*25*, *34*). The reduced resistivity in the thinner NbP/Nb films that was maintained down to ~5 K suggested a non-negligible contribution of surface conduction in these samples (*29*, *35*, *36*).

To obtain the conductance of the NbP layer (**Fig. 3D)**, we subtracted the sheet conductance of the 4-nm-Nb seed layer (**fig. S6C**) from the total sheet conductance of the NbP/Nb stack (**Fig. 3C**) over the 5 to 300 K temperature range. The extracted resistivity of the NbP layer also showed the unconventional trend of decreasing resistivity with decreasing NbP thicknesses from room temperature down to 5 K (**Fig. 3E**). To better understand the trend in **Fig. 3E** quantitatively, we fit the conductance of the NbP layers (with various thicknesses) with both bulk and surface channel contributions to the conductance between 5 and 300 K (**Fig. 3F, fig. S12** and **fig. S13**). We assumed that the NbP surface conductance contribution was constant with the sample thickness (further details in Supplementary **Text Section I: Surface and Bulk Conductance of NbP/Nb and NbP Layer** and **fig. S12**).

As can be seen from **Fig. 3F** (and **fig. S13**), the bulk conductance of our NbP films increased from 5 to 300 K, as expected for variable-range hopping behavior in amorphous and nanocrystalline films (*25*). In contrast, the surface conductance was metallic and decreased with increasing temperature (*25*, *29*). As thickness decreased from ~80 to ~4.3 nm, the bulk contribution to the conductance decreased. At low temperatures, we expected the hopping carrier transport to be small and nearly independent of sample thickness. Thus, the conduction was dominated by a surface channel at low temperature (e.g., < 50 K) even in the thicker 80-nm NbP sample (*25*, *29*).

We can also estimate the surface-to-bulk conductance ratio in **Fig 3G**, which revealed that all the thinner films (18 nm NbP or less) were dominated by their surface contribution up to room temperature. The resistivity of our 4.3 nm NbP film was smaller than the bulk single-crystal NbP resistivity (*14*, *32*), whereas the resistivity of our 80 nm NbP film was ~3× higher than the single-crystal value. The lower resistivity of our thinner NbP was unlikely the result of improved crystallinity because these films were predominantly amorphous with embedded nano-crystallites (**Fig. 2, A and B**) and had higher conductivity than bulk single-crystal NbP.

We also estimated the bulk NbP conductance and the effective surface conductance of NbP (with the Nb layer) from the total sheet conductance of the NbP/Nb samples vs. NbP thickness in **fig. S14**, with the analysis detailed in **Materials and Methods**. Supplementary **fig. S14A** shows that the surface conductance dominated the total sheet conductance for all NbP/Nb film stacks thinner than ~30 nm at room temperature. Even in the presence of defects or disorder, the higher conductivity in our thinner NbP/Nb films and NbP layers came from a surface-like channel.

**Carrier density estimates**

We performed Hall resistance measurements of our NbP films as a function of magnetic field at 5 K (**Fig. 4A**). We subtracted the deduced Hall conductivity of the 4-nm-Nb seed layer (**fig. S15A**) from that obtained for our stacks (**fig. S15B**). From **Fig. 4A**, the Hall resistance was linear with magnetic field at all sample thicknesses, suggesting a single carrier dominated transport in our NbP films (in this case, holes). The Hall resistance of our 4.3-nm-thick NbP versus magnetic field was nearly independent of temperature between 5 and 20 K (**Fig. 4B**). The extracted sheet carrier density at 5 K in **Fig. 4C** decreased from ~$10^{18}$ $cm^{-2}$ for 80 nm thick NbP to ~$10^{16}$ $cm^{-2}$ in 4.3 nm thin NbP (see details in **Supplementary Text**). This trend was consistent with previous reports on thicker films of crystalline topological semimetals NbAs and TaAs (*23*, *35*).

We note that the carrier density per unit volume in our NbP films (> $10^{22}$ $cm^{-3}$ in **fig. S16**) was higher (*45*) than in NbP bulk single crystals (*14*) but comparable to other topological semimetals such as ~70 nm thick NbP epitaxial films (> $10^{22}$ $cm^{-3}$) (*32*), textured and amorphous CoSi (*46*), and topological metals like MoP (> $10^{23}$ $cm^{-3}$) (*30*). We also note that effective carrier density estimated from Hall measurements in disordered or non-crystalline films, like our NbP, could be overestimated (and the mobility underestimated) due to possible contribution from hopping-like transport (*47*). This has been reported in organic semiconductors (*47*) and the topological insulator $Bi_2Se_3$, where the total carrier density estimated in non-crystalline films was ~10 times higher (*25*) than in its crystalline counterpart (*29*).

The carrier density vs. thickness trend (**Fig. 4C**) allows us to estimate an average surface carrier density of ~$10^{16}$ $cm^{-2}$, i.e. the hole density in the limit of the NbP film thickness approaching zero; this projected surface carrier density in our non-crystalline NbP is ~3 times larger than what was estimated in crystalline NbAs (*23*); however, it is consistent with the possibility of a higher apparent carrier density from Hall measurements in a non-crystalline system, as explained above.

The estimated mobility at 5 K (**Fig. 4D**) shows an increasing trend with decreasing NbP thickness. The effective mobility (at 5 K) of a 4.3 nm thin NbP film is ~7.4 $cm^2V^{-1}s^{-1}$, approximately 50 times greater than that of the 80 nm thick NbP film (~0.15 $cm^2V^{-1}s^{-1}$). Using the extrapolated surface sheet carrier density (**Fig. 4C**) and surface conductance (**Fig. 3F**), we estimate the mobility (see **Supplementary Text**) of the surface-like channel to be 9.4 ± 3.0 $cm^2V^{-1}s^{-1}$. This higher surface mobility appears to enable the lower resistivity in our thinnest NbP films (**Fig. 3E**), where conduction is dominated by surface rather than bulk channels (**Fig. 3G**). We recall that these estimates were performed after careful subtraction of the 4 nm Nb seed layer contribution (**fig. S6**); however, we find that the thickness-dependent carrier density and mobility trends shown in **Fig. 4C,D** are maintained even when the Nb layer is included, i.e., in NbP/Nb heterostructures (see **fig. S18**).

What are the origins of the surface-like conduction in these ultrathin nanocrystalline films? This remains a partly open question, but we suggest a few possible causes. One possibility is the formation of disorder-tolerant Fermi arc-like surface states (*23*) even in non-crystalline topological materials (*45*, *48*); another cause may be the existence of an interfacial free-electron gas-like state (*29*) near the NbP/Nb interface, where we observed local short-range ordering (**Fig. 2A,B** and **fig. S4**). For example, topological surface states are expected to be metallic-like in nature (*25*) and less sensitive to disorder scattering (*19*, *23*). The estimated surface mobility (~9.4 $cm^2V^{-1}s^{-1}$ at 5 K) of our non-crystalline NbP films is much lower than that of crystalline NbP (~$10^6$ $cm^2V^{-1}s^{-1}$ at ~2 K) (*14*) and topological insulators such as $Bi_2Se_3$ (~$10^3$ $cm^2V^{-1}s^{-1}$ at 1.5 K) (*29*). However, the surface mobility in our films is comparable with mobilities found in sub-10 nm thin polycrystalline $Bi_2Se_3$ (< 10 $cm^2V^{-1}s^{-1}$ at 1.5 K) (*29*, *36*) and in thick amorphous $Bi_2Se_3$ (< 20 $cm^2V^{-1}s^{-1}$ at 2 K) (*25*) with topological surface states. In the end, the low resistivity of our ultrathin NbP films is caused by the combination of high surface carrier density (~$10^{16}$ $cm^{-2}$) and sufficiently good surface mobility. We also recall that the low resistivity is surface-dominated and maintained up to room temperature in all sub-18 nm thin films (**Fig. 3**). Looking ahead, we expect our work to motivate future efforts into imaging surface state dispersion in amorphous or non-crystalline semimetals like NbP using surface-sensitive techniques such as angle-resolved photoemission spectroscopy (ARPES) and spin-resolved ARPES (*25*).

In summary, we uncovered that the resistivity of amorphous or nanocrystalline films of NbP decreases dramatically as the film thickness is reduced, which is a trend counter to that observed in most common metals. The thinnest films (< 5 nm) display resistivities lower than conventional metals of similar thickness, at room temperature. Measurements and modeling indicate that our NbP films thinner than ~18 nm are dominated by surface conduction up to room temperature, which is the origin of the effective resistivity decrease in thinner films. Importantly, these films were deposited by large-area sputtering at relatively low temperatures (400 °C), compatible with modern microelectronics processing. These results and the fundamental insights obtained here could enable ultrathin topological semimetals as low-resistivity interconnects in future high-density electronics and spintronics.

## Acknowledgements

We dedicate this work in memory of the late Prof. AKM. Newaz (San Francisco State University) and Prof. Evan Reed (Stanford University). We are grateful for discussions with them throughout the years, and for their contribution to the scientific community. Authors also acknowledge Prof. H.-S. Philip Wong for discussions and encouragement related to this work. A.I.K. is thankful to James McVittie and Carsen Kline for their support and discussion regarding materials deposition, to Swaroop Komera for the lab support, and to Christian Lavoie for additional insights. The TEM work was supported by Taehoon Cheon at the Daegu Gyeongbuk Institute of Science and Technology (DGIST). A.I.K. also thanks Maliha Noshin and Heungdong Kwon for useful discussions on materials deposition.

**Funding:** This work was performed at the Stanford Nanofabrication Facility (SNF) and Stanford Nano Shared Facilities (SNSF), supported by the National Science Foundation (NSF) award ECCS-2026822. The Stanford authors were supported in part by the Precourt Institute for Energy and the SystemX Alliance; A.I.K acknowledges support from the Stanford Graduate Fellowship. E.L. and Y.S. were funded by the NSF on Award 2037652. A.R. and F.H.J. acknowledge support from the NSF program Designing Materials to Revolutionize and Engineer our Future (DMREF) project DMR-1922312. I-K.O. acknowledges support from the National Research Foundation of Korea (NRF) grant funded by the Korea government (MSIT) (RS-2024-00357895).

**Author contributions:** A.I.K. conceived the idea with C.T.C, supported by E.P and K.S. A.I.K. formulated and optimized the material deposition process with input from C.T.C. and K.S. A.I.K. led the design of experiments with input from E.P. A.I.K performed the materials deposition with help from X.W. and H.K, and inputs from E.P. TEM, EDS, XPS and strain analysis were performed by H-M.K. I-K.O. and B.W. with lead and input from I-K.O. A.I.K fabricated the devices for transport measurements. E.L. performed the Hall measurements with input from Y.S., A.I.K., and A.R. A.R. The transport data were analyzed by A.R. with input from A.I.K., F.H.J., and E.P. A.I.K., A.R., I-K.O., and E.P. wrote the manuscript, with input from all authors.

**Competing interests:** The authors declare no competing interests. **Data and materials availability:** All data needed to evaluate the conclusions in this paper are present in the paper or the supplementary materials.

## Supplementary Materials

Materials and Methods

Supplementary Text (Sections I and II)

Figs. S1 to S18

Table S1

References

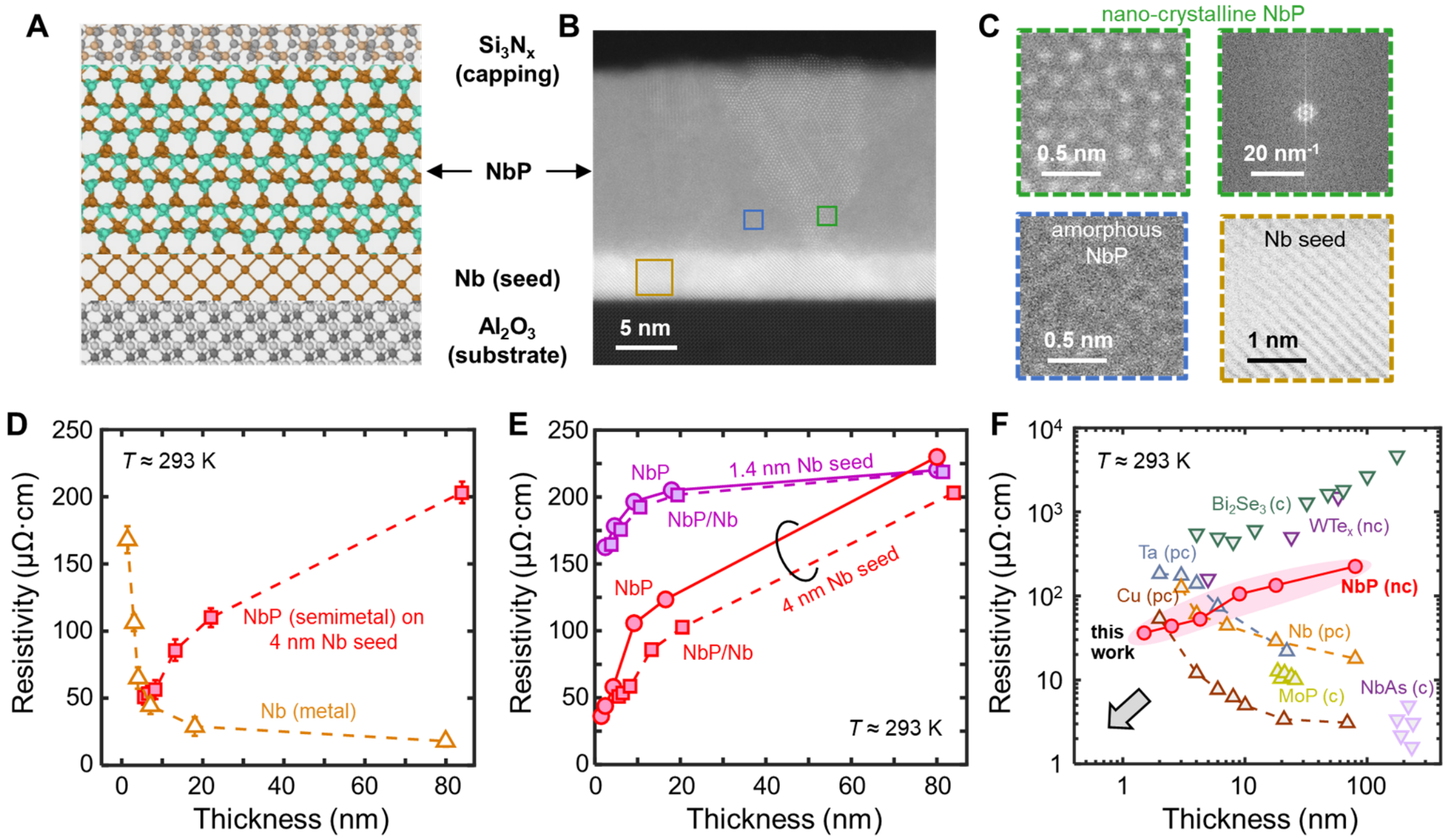

**Fig. 1. NbP/Nb thin film stacks and room temperature resistivity**. **(A)** Schematic of the sputtered NbP/Nb film stack. **(B)** High-angle annular dark-field (HAADF) scanning transmission electron microscope (STEM) cross-section of an ~18-nm NbP/Nb film stack. The $Al_2O_3$ substrate, thin (~4-nm) Nb seed layer, and silicon nitride capping can also be seen. **(C)** Zoomed-in STEM images show local short-range ordering and nanocrystallinity within the amorphous NbP layer, as well as the crystallinity of the Nb seed layer (see **fig. S2** for other NbP thicknesses). **(D)** Room-temperature resistivity versus thickness of NbP/Nb films (squares), and of control Nb films (triangles). The resistivity and thickness of NbP plotted here includes the 4-nm Nb seed layer. The NbP/Nb stack showed unconventional resistivity scaling in that the resistivity decreased in thinner films. Symbols and error bars mark the average and standard deviation, respectively, across five samples of each film thickness. **(E)** Room-temperature resistivity versus thickness of NbP/Nb stacks before (squares) and after (circles) subtracting the Nb seed-layer conduction contribution. Samples with two different Nb seed layers are shown: 4 nm (red) and 1.4 nm (violet). Unconventional resistivity scaling is noted for all films, both including and excluding the Nb seed layer contribution. The horizontal axis represents either the total stack thickness (NbP + Nb) or just the NbP thickness. **(F)** Room-temperature resistivity versus thickness for various materials. Here, our sputtered NbP semimetal resistivity is shown after subtracting the contribution of the Nb seed; similarly, Cu resistivity is shown without the contribution of its liner and barrier layers (*28*). Other films include Nb (from this work), Ta, crystalline topological insulator $Bi_2Se_3$ (*29*), nanocrystalline topological semimetals $WTe_x$ and NbAs (*23*, *26*), and topological metal MoP (*30*). The arrow marks the best corner region of smallest resistivity at low film thickness. (c), (pc) and (nc) refer to crystalline, polycrystalline and noncrystalline films, respectively. Our sputtered NbP displayed decreasing resistivity down to sub-5 nm thickness, with the lowest resistivity in ultrathin films.

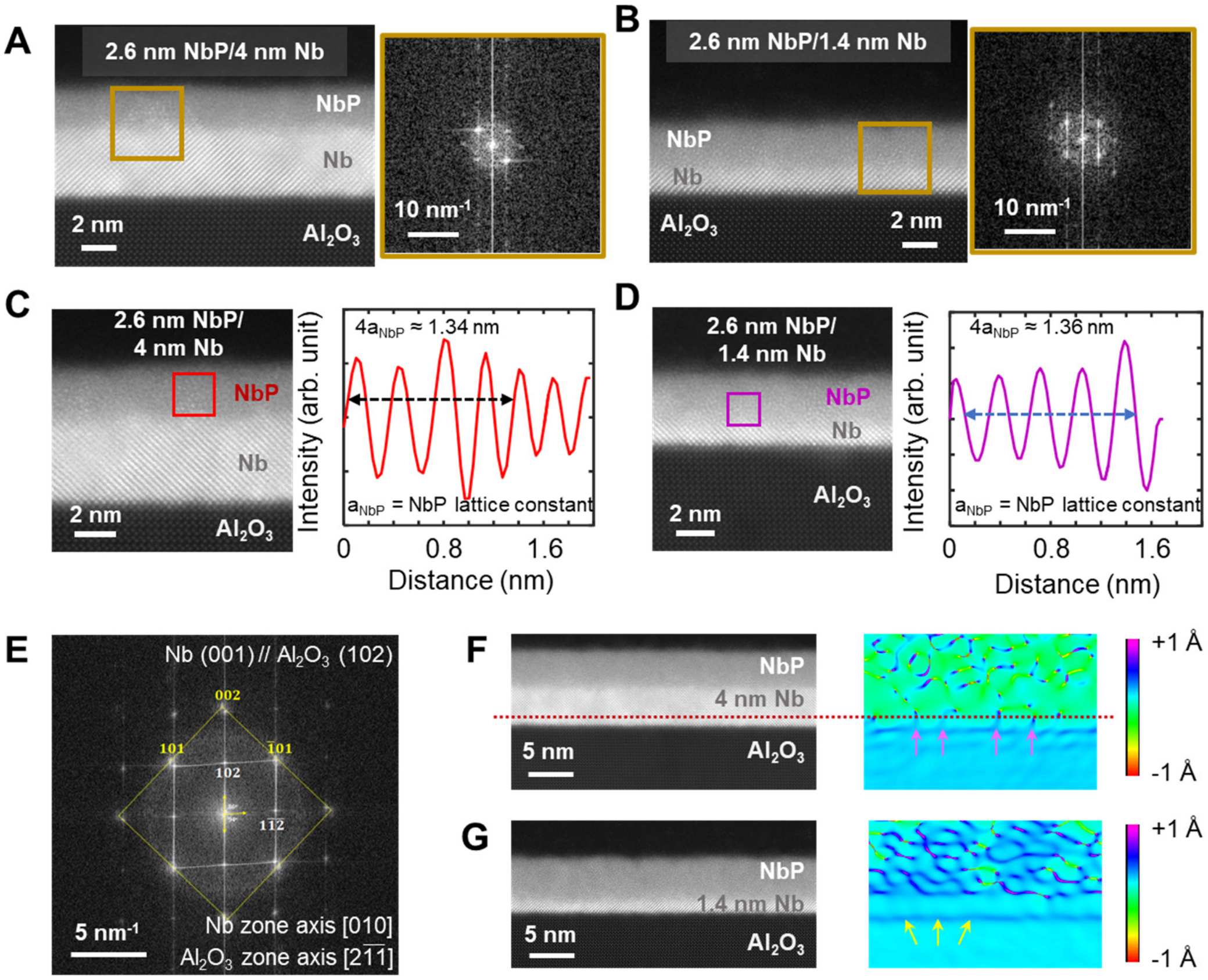

**Fig. 2. Microstructure details of ultrathin NbP/Nb heterostructures. (A and B)** HAADF-STEM images and their fast Fourier transform (FFT) diffraction of 2.6-nm NbP films on **(A)** Nb seed layer with 4- nm and **(B)** 1.4-nm thickness. Local nanocrystalline (short-range order) region of 2.6-nm thin NbP film on **(C)** 4-nm Nb seed, showing NbP lattice constant ~3.34 Å, near its nominal value of ~3.332 Å (*42*). **(D)** Similar image on 1.4 nm Nb seed layer, revealing NbP lattice constant ~3.41 Å, which indicated that NbP was strained on the thinner Nb seed. **(E)** Diffraction pattern of Nb seed layer and $Al_2O_3$ substrate. Nb seed layers have an epitaxial relationship with the $Al_2O_3$ substrate – Nb (001) ǁ $Al_2O_3$ (102). The Al in $Al_2O_3$ (102) has a rhombus lattice tilted by 6° compared to Nb (001). **(F)** Lattice strain analysis of 2.6-nm NbP film on 4-nm Nb and **(G)** on 1.4-nm Nb from Fourier filtering the corresponding HAADF-STEM images. The 1.4-nm Nb seed was strained laterally along the $Al_2O_3$ surface (yellow arrows), but the accumulated strain was released in the 4-nm Nb seed by forming misfit dislocations (pink arrows); the red dotted line marks the level of dislocations within the Nb seed. The colored images display the strain mapping of the layers. The greater green proportion in the top plot marks a larger unstrained portion of NbP on the thicker (~4-nm) Nb seed, compared to the thinner (~1.4-nm) one.

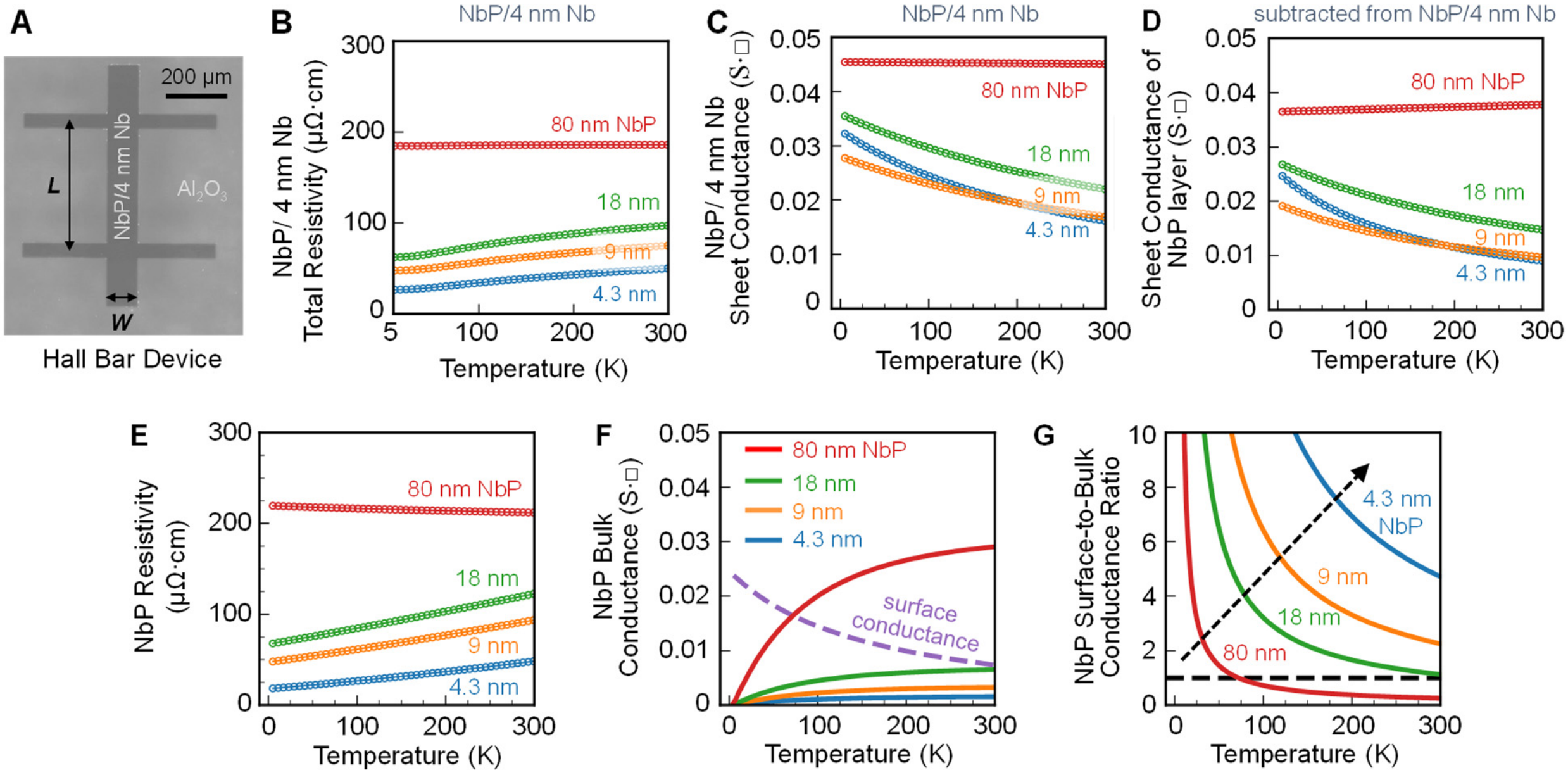

**Fig. 3. Temperature-dependent transport of NbP/Nb and NbP**. **(A)** Top view optical image of the Hall bar with width $W$ of 100 μm and length $L$ of 400 μm. The NbP was seeded by Nb and capped by $SiN_x$, as in Fig. 1B. **(B)** Temperature-dependent total resistivity of NbP/Nb, and **(C)** sheet conductance of NbP/Nb films with varying NbP thicknesses (here 4.3, 9, 18, and 80 nm) on a 4-nm Nb seed. **(D)** Sheet conductance of the NbP layer of varying thicknesses obtained by subtracting the conductance of the 4-nm Nb seed layer (measured separately, fig. S4A) from the total sheet conductance of NbP/Nb films in **C**. **(E)** Temperature-dependent resistivity of NbP films with varying thicknesses, from 4.3 and 80 nm (obtained using **D**). **(F)** Two-channel conductance fit to the resistivity data in **E**, indicating a metallic surface channel (dashed line) and disorder-dominated bulk conductance (solid lines). Here, we assumed the surface channel has zero thickness. Figure S13B displays the fit with a finite surface thickness ≈ 5 Å, yielding a similar result. **(G)** Calculated surface-to-bulk conductance ratio versus temperature for our NbP films. The surface-to-bulk conductance ratio increased as the NbP film thickness is reduced (indicated by the dashed black arrow) across a wide range of temperatures. The region above the dashed line was dominated by surface conduction.

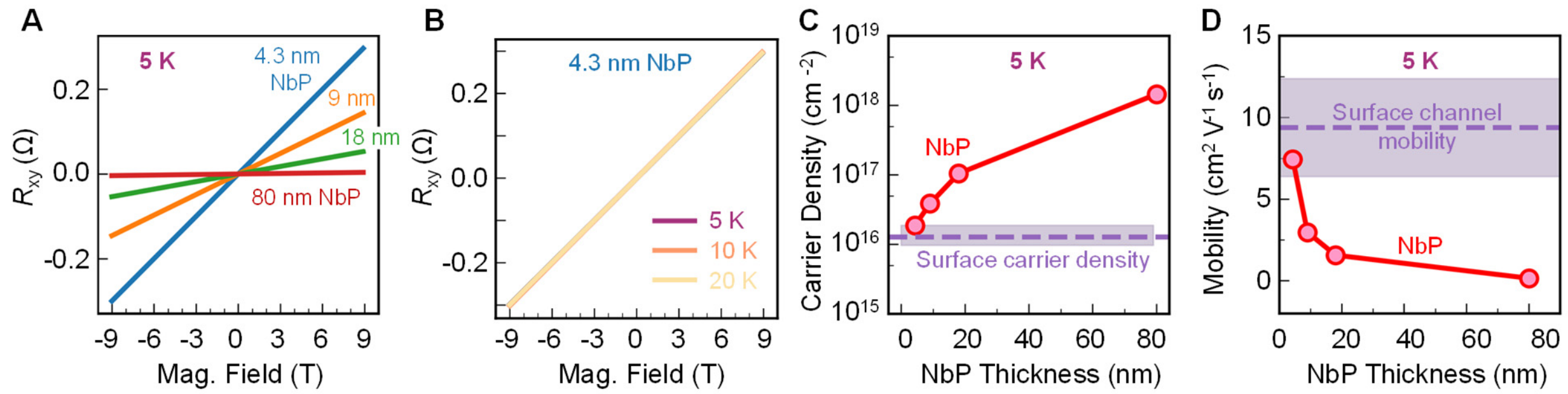

**Fig. 4. Hall measurements and carrier densities of our NbP films.** **(A)** Hall resistance vs. magnetic field for NbP films with varying thicknesses at 5 K. **(B)** Hall resistance of a 4.3 nm thin NbP film vs. magnetic field at 5 K, 10 K, and 20 K. **(C)** Sheet carrier density (holes, extracted from Fig. 4A) shows reduction with NbP film thickness. From the Hall coefficient vs. thickness fit in **fig. S17** we estimate a surface carrier density of $1.4 \pm 0.4 \times 10^{16}$ cm$^{-2}$, the sheet carrier density in the limit of zero NbP film thickness; shaded purple region. **(D)** Mobility of the NbP films, showing an increasing trend with decreasing thicknesses. The shaded region represents the range of the surface channel mobility, 9.4 ± 3.0 cm$^2$V$^{-1}$s$^{-1}$, estimated from the surface carrier density. All data and estimates in this figure are after subtracting the conduction contribution of the 4 nm Nb seed (see Supplementary **Materials and Methods** and **fig. S15**) We note that including the conduction contribution of the 4 nm Nb seed layer does not alter the carrier density and mobility trends shown in **C, D** (see **fig. S18**).

# Supplementary Materials for

## Surface conduction and reduced electrical resistivity in ultrathin noncrystalline NbP semimetal

Asir Intisar Khan, Akash Ramdas, Emily Lindgren, Hyun-Mi Kim, Byoungjun Won, Xiangjin Wu, Krishna Saraswat, Ching-Tzu Chen, Yuri Suzuki, Felipe H. da Jornada, Il-Kwon Oh*, Eric Pop*

*Corresponding author: epop@stanford.edu, ikoh@ajou.ac.kr

**The PDF file includes:**

## Materials and Methods

### Materials Deposition

In this work, we prepared four types of film samples:

1) **NbP/Nb films** on insulating r-plane sapphire ($Al_2O_3$) or MgO substrates. These were sputter-deposited at 400 °C (**fig. S1, table S1**), a temperature compatible with back-end-of-the-line semiconductor fabrication. Direct current (dc) magnetron sputtering was performed at 20 W power and 3 mTorr pressure. To reduce lattice mismatch between the substrate and NbP, we first deposited a thin buffer (seed) layer of Nb between 1.4 to 4 nm thickness (*29*). Then, the NbP film was deposited, ranging from 1.5 nm to 80 nm thickness, at a rate of 1.1 nm/min. The Nb seed and NbP deposition were at 400 °C, a temperature which was optimized (in the 300 to 800 °C range) to produce films with lowest resistivity.
2) **NbP/Nb films** on $SiO_2$ (amorphous) on Si substrates. The NbP thickness was 2.6 nm and 4.3 nm, the Nb seed thickness was 4 nm, and depositions conditions were as stated above.
3) **Cu/Nb films** with 4 nm Nb seed layer, on r-plane sapphire. The Nb seed was deposited as stated above, and Cu metal films (2.5 nm to 20 nm thick) were sputtered at room temperature.
4) **Nb films** on r-plane sapphire with the same thickness as the Nb seed layers used for NbP.

All film samples in this work were capped with 3 to 4 nm thin $SiN_x$ layer, deposited at room temperature, to prevent surface oxidation. All layers were deposited without breaking vacuum.

### Materials Characterization

We used a double spherical aberration (Cs) corrected transmission electron microscopy (Themis Z, ThermoFisher Scientific) with an 80 pm resolution and an acceleration voltage of 200 kV. For the atomic-resolution imaging with high angle annular dark field (HAADF) scanning transmission electron microscopy (STEM), we used a probe convergence angle of 15.3 mrad and the inner collection semi angles of 70 mrad and 200 mrad. Energy dispersive spectroscopy (EDS) with four windowless detectors (SuperXG2) was used for the composition mapping of our NbP samples. X-ray photoelectron spectroscopy (Thermo Fisher Scientific, NEXSA) was performed with 400 $\mu m^2$ of X-ray spot size and 1000 eV of ion gun energy.

## Device Fabrication and Electrical Measurement

After film deposition, the substrates were cut into rectangular shapes (7.5 × 8.5 $mm^2$). On these, we patterned standard Hall bar devices (**Fig. 3A**) using direct-write lithography (Heidelberg MLA 150) followed by reactive ion etching of the $SiN_x$/NbP/Nb stack. For the reactive ion etching, we used 30 sccm $Cl_2$ / 10 sccm $CH_4$, 60 W RF power at a pressure of 10 mTorr. Contacts were directly wirebonded (punching through the thin $SiN_x$ capping layer) to the top and side Hall bar edges. All temperature-dependent electrical transport measurements (5 to 300 K) were performed under vacuum in a Quantum Design Dynacool system, using the geometry shown in Fig. 3a. Magnetoresistance measurements used magnetic fields up to ± 9 T in the out-of-plane direction. Additional room temperature electrical resistivity measurements (Fig. 1d,e) were performed in a Lakeshore 8404 Hall measurement system and an LEI1500 Eddy current system.

## Supplementary Text

### Section I: Surface and Bulk Conductance of NbP/Nb and NbP Layer

We can write the total sheet conductance, $G$ (in S·□) of our NbP/Nb films at temperature $T$ as:

$$G(t_{\mathrm{Nb}}, t_{\mathrm{b}}, T) = G_{\mathrm{Nb}}(t_{\mathrm{Nb}}, T) + \underbrace{G_{\mathrm{b}}(t_{\mathrm{b}}, T) + G_{\mathrm{s}}(T)}_{G_{\mathrm{NbP}}} \tag{1}$$

Where $G_{Nb}$ is the Nb seed layer conductance, $G_b$ is the bulk NbP conductance, $G_s$ is the NbP surface conductance, $t_{Nb}$ and $t_b$ are the Nb seed layer and NbP film thickness, respectively. Here, $G_b = \sigma_b(T)t_b$, is the product of the bulk NbP conductivity and the NbP thickness.

From the total sheet conductance $G$ of NbP/Nb films with varying thicknesses (**Fig. 3C** and **fig. S6D**), we can extract the bulk and surface conductance contributions of the NbP/Nb heterostructure at different temperatures (**fig. S14)** by rearranging eq. (1):

$$G(t_{\mathrm{Nb}}, t_{\mathrm{b}}, T) = G_{\mathrm{b}}(t_{\mathrm{b}}, T) + \underbrace{G_{\mathrm{Nb}}(t_{\mathrm{Nb}}, T) + G_{\mathrm{s}}(T)}_{G_{\mathrm{s,NbP/Nb}}} \tag{2}$$

Where, the extracted $G_{s,NbP/Nb}$ is the 'effective' surface conductance which includes the conduction contribution of the bottom NbP surface and the 4 nm Nb seed layer.

Next, we calculated the conductance of the NbP layer (**Fig. 3D**) by subtracting the conductance of the Nb seed layer (measured separately, **fig. S6C)** from the total sheet conductance, $G$ of NbP/Nb films (**Fig. 3C**) using $G_{NbP} = G(t_{Nb}, t_b, T) - G_{Nb}(t_{Nb}, T)$. We then extract the $T$-dependent resistivity ($\rho_b = t_b/G_{NbP}$) of four different thicknesses of NbP ($t_b \approx 4.3$, 9, 18, 80 nm) (**Fig. 3E**) and subsequently estimated $\sigma_b(T)$ and $G_s(T)$ from $G_{NbP} = \sigma_b(T)t_b + G_s(T)$ (**fig. S12A**, **Fig. 3F, fig. S13A**).

Instead of an ideal two-dimensional surface with thickness $t_s = 0$ Å (as in **fig. S12A**), if we assume that NbP has a finite surface thickness $t_s = 5$ Å (**fig. S12B)**, the total conductance of NbP is $G_{NbP} = \sigma_b(T)(t_b - t_{s,}) + \sigma_s(T)t_s$. In this case, the estimated bulk and surface conductance of four different thicknesses of NbP films are shown in **fig. S13B**, between 5 K to 300 K. Here, the bulk conductance increases with temperature from 5 K to 300 K. In contrast, the surface conductance decreases with increasing temperature, and as the NbP is thinned from 80 nm to 4.3 nm, the bulk channel contribution to the conductance decreases for thinner films.

**Section II: Carrier Density and Mobility Estimation**

We estimate an effective sheet carrier density (in cm$^{-2}$) $n = 1/(qR_H)$, where $q$ is the elementary charge and $R_H$ is the Hall coefficient (slope of the transverse Hall resistance $R_{xy}$ vs. magnetic field $B$ at 5 K temperature, in **Fig. 4A**) shown in **fig. S17**. From the estimated carrier density $n$ (here, holes) and the sheet conductance of NbP ($G_{NbP}$) we obtain an effective mobility $\mu = G_{NbP}/(qn)$.

The longitudinal sheet conductance $G_{NbP} = \sigma_b t_b + G_s$ and the transverse (Hall) conductance $G_{xy} = (\sigma_b t_b + G_s)^2/[B(\sigma_b^2 t_b^2 R_{H,b} + G_s^2 R_{H,s})]$, where $R_{H,s}$ and $R_{H,b}$ are Hall coefficients of the surface and bulk charge carriers (holes), respectively. We can rearrange this expression as $R_H = 1/(qn) = 1/(BG_{xy}) = R_{H,b}\,(\sigma_b t_b)^2(\sigma_b t_b + G_s)^{-2} + R_{H,s}\,(G_s)^2(\sigma_b t_b + G_s)^{-2}$. When the film thickness approaches zero, we can write $R_H(t_b \rightarrow 0) = R_{H,b} \times 0 + R_{H,s} \times 1 = R_{H,s}$. In other words, as $t_b$ approaches zero, $BG_s = 1/R_{H,s}$. Then, from the measured Hall coefficient $R_H$ vs. NbP thickness, we can estimate $R_{H,s}$ by finding the $t_b \rightarrow 0$ limit of $R_H$ (**fig. S17)**. To estimate the uncertainty of this approach, we used the measured $R_H$ of our thinnest NbP film (here 4.3 nm) as a lower bound for $R_{H,s}$.

We can estimate the surface mobility, $\mu_s = G_s/(qn_s)$, where $n_s = 1/(qR_{H,s})$. To extract the carrier density and mobility of NbP, we subtract the Hall conductance of the 4 nm Nb seed layer (**fig. S15A**) from that of the NbP/Nb film stacks (**fig. S15B**), using $G_{xy,NbP} = G_{xy} - G_{xy,Nb}$, where $G_{xy}$ is the measured total Hall conductance of the NbP/4 nm Nb film, and $G_{xy,Nb}$ is the Hall conductance of our reference 4 nm Nb seed film. We note that the Hall measurement of our 4 nm Nb film (fig. S15A) yields $2.17 \times 10^{23}$ $cm^{-3}$ volumetric electron density. This is comparable but higher than the carrier density obtained in (*49*). The Nb seed layer also shows superconducting behavior below 2.5 K (Fig. S6), as expected in such thin films (*49, 50*).

We repeated our transport analysis of Fig. 4 in the main text *without* subtracting the contribution of 4 nm Nb seed, as shown in **fig. S18**. Here, we find that even for the NbP/Nb heterostructures, the transport *trends* described in Fig. 4 from the main text (e.g., carrier density, mobility) remain unchanged. NbP/Nb heterostructures (including 4 nm Nb seed) also show a decreasing carrier density, and an increasing mobility with decreasing total stack thickness.

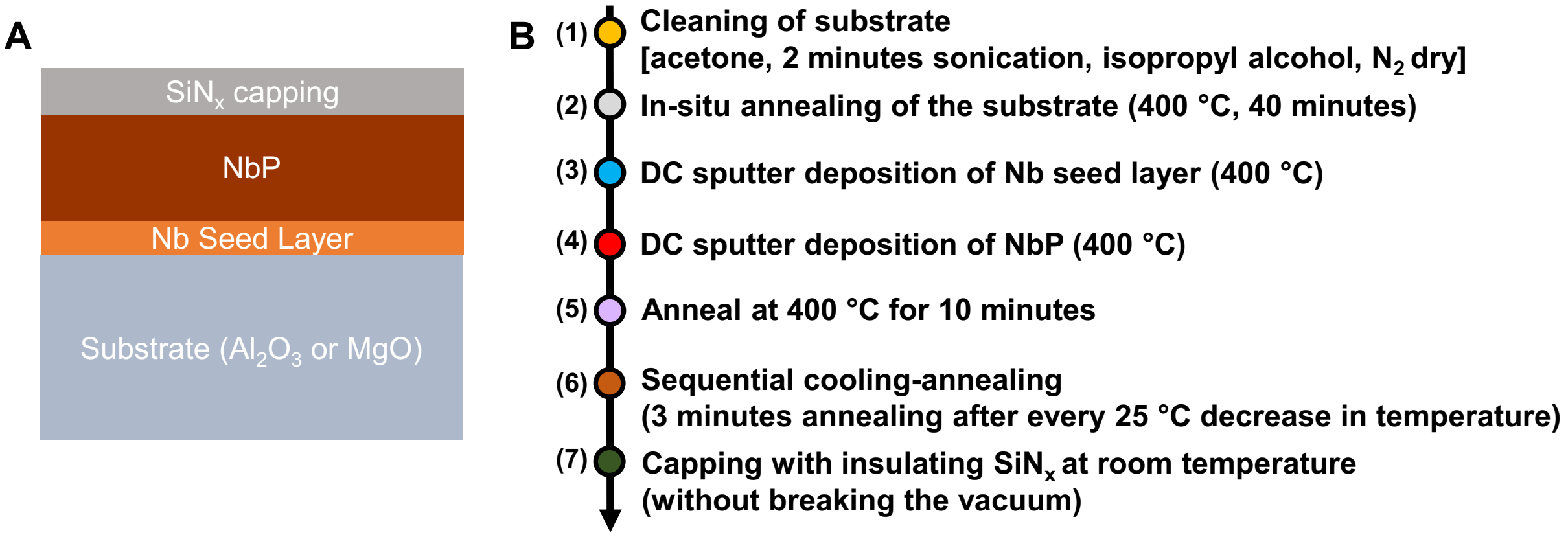

**Fig. S1. Materials deposition steps. (A)** Schematic of the NbP semimetal stack on top of a thin Nb seed layer. **(B)** Sputtering steps which form the NbP thin film stacks. The chamber base pressure was kept below $5 \times 10^{-8}$ Torr. See **Materials and Methods: Materials Deposition** section for additional details.

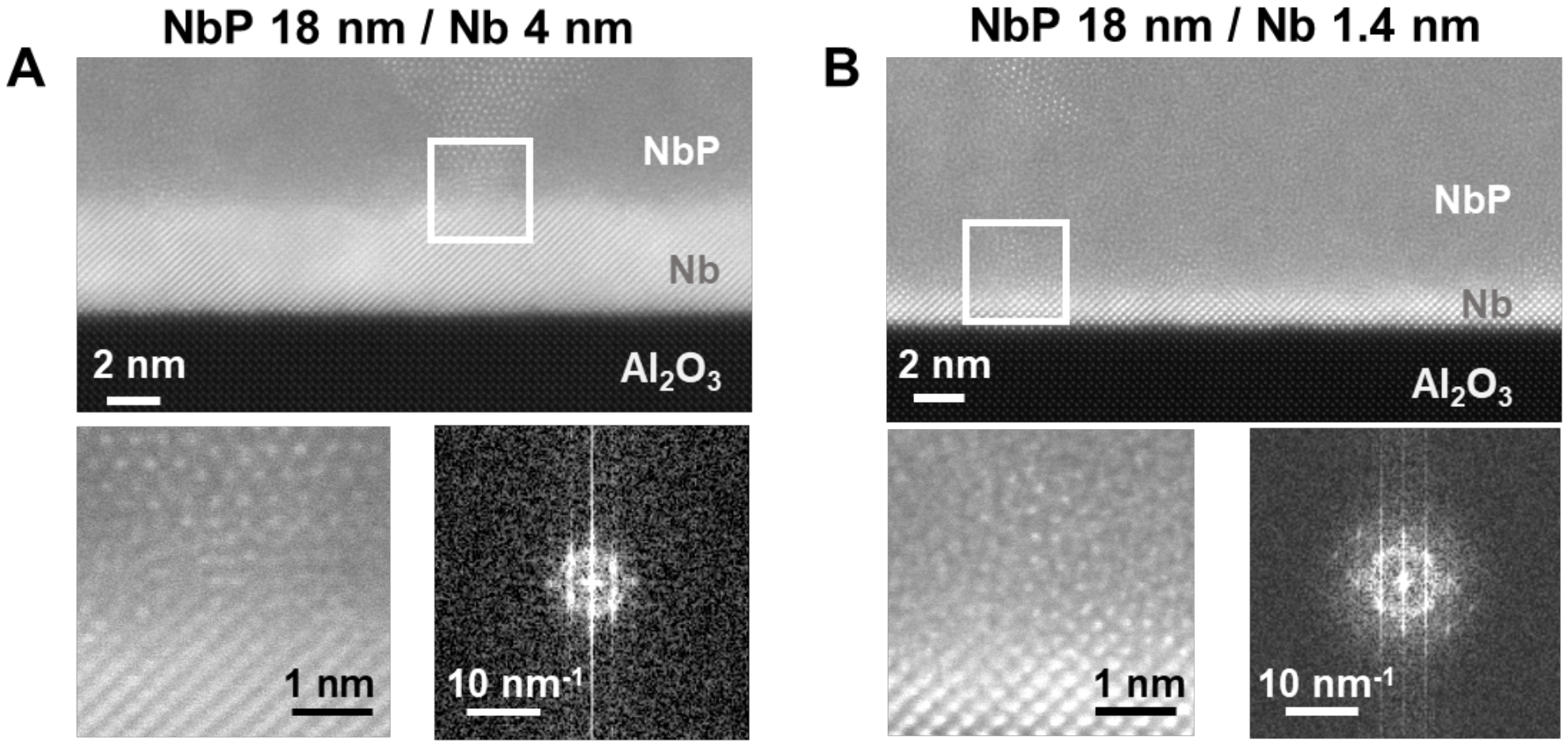

**Fig. S2. STEM and diffraction patterns of NbP films.** High resolution HAADF-STEM and zoomed-in images and corresponding diffraction patterns for **(A)** 18 nm NbP film on a 4 nm Nb, and **(B)** 18 nm NbP on a 1.4 nm Nb seed layer, showing a similar nano-crystallinity of the NbP films near the NbP/Nb interface for both 4 nm and 1.4 nm Nb seed layers. We note that the Nb seed layer shows a comparable crystalline quality as the control films (**fig. S3**), and its resistivity is much higher in the thinnest films (see Fig. 1D) compared to that of the NbP/Nb stack.

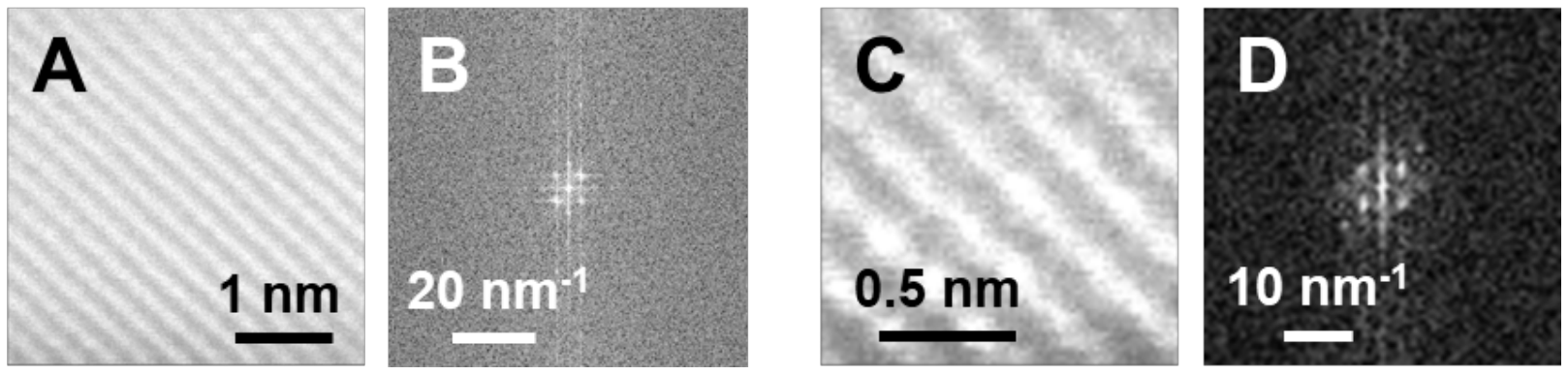

**Fig. S3. STEM and diffraction patterns of control thin Nb films.** High-resolution HAADF-STEM and diffraction patterns of **(A,B)** a 4 nm Nb, and **(C,D)** a 1.4 nm Nb film showing a similar degree of crystallinity compared to the Nb seed layers in the NbP/Nb stack.

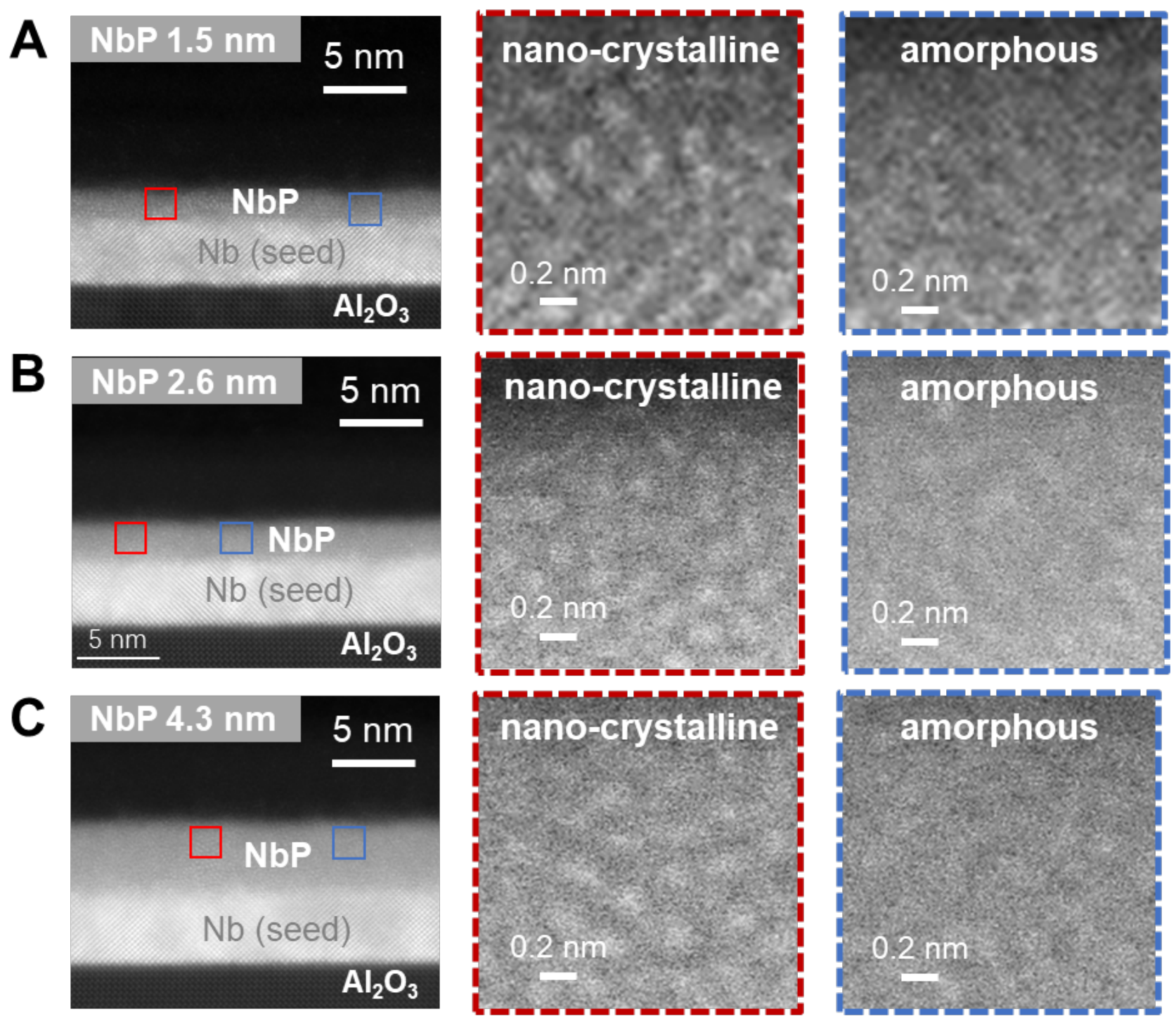

**Fig. S4**. **STEM characterization of NbP films on 4 nm Nb seed**. Zoomed-in HAADF-STEM images of **(A)** 1.5 nm, **(B)** 2.6 nm, and **(C)** 4.3 nm NbP films on a 4 nm Nb seed layer showing the presence of local short-range ordering and nano-crystallinity (red-box panels) within the amorphous NbP film matrices. Red and blue box panels display representative nano-crystalline and amorphous regions, respectively.

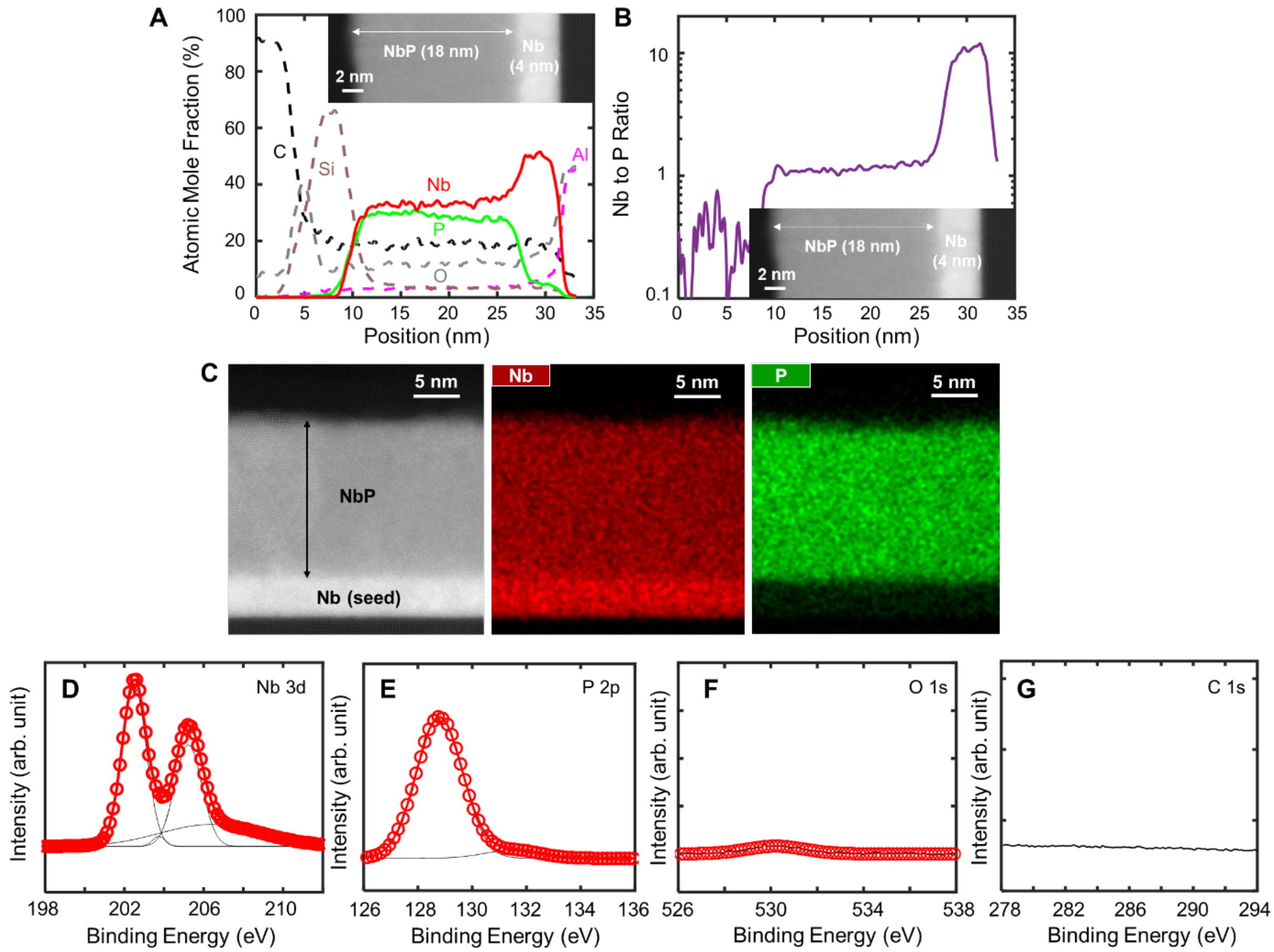

**Fig. S5. EDS and XPS characterization of NbP/Nb film. (A,B)** Energy dispersive spectra (EDS) line scans showing the atomic ratio between Nb and P in our NbP film (here ~18 nm) to be close to 1. **(C)** EDS compositional mapping performed from HAADF-STEM confirming the homogeneity of Nb and P across the NbP sample. The presence of C and O elements in EDS characterization could be due to hydrocarbons adsorbed onto the sample surface, and surface oxidation during sample preparation. XPS spectra of an 18 nm NbP/4 nm Nb film (after 120 s etching): **(D)** Nb 3d, **(E)** P 2p, **(F)** O 1s, and **(G)** C 1s core levels, display no significant carbon incorporation and a small percentage (~4 %) of O inside the NbP layer.

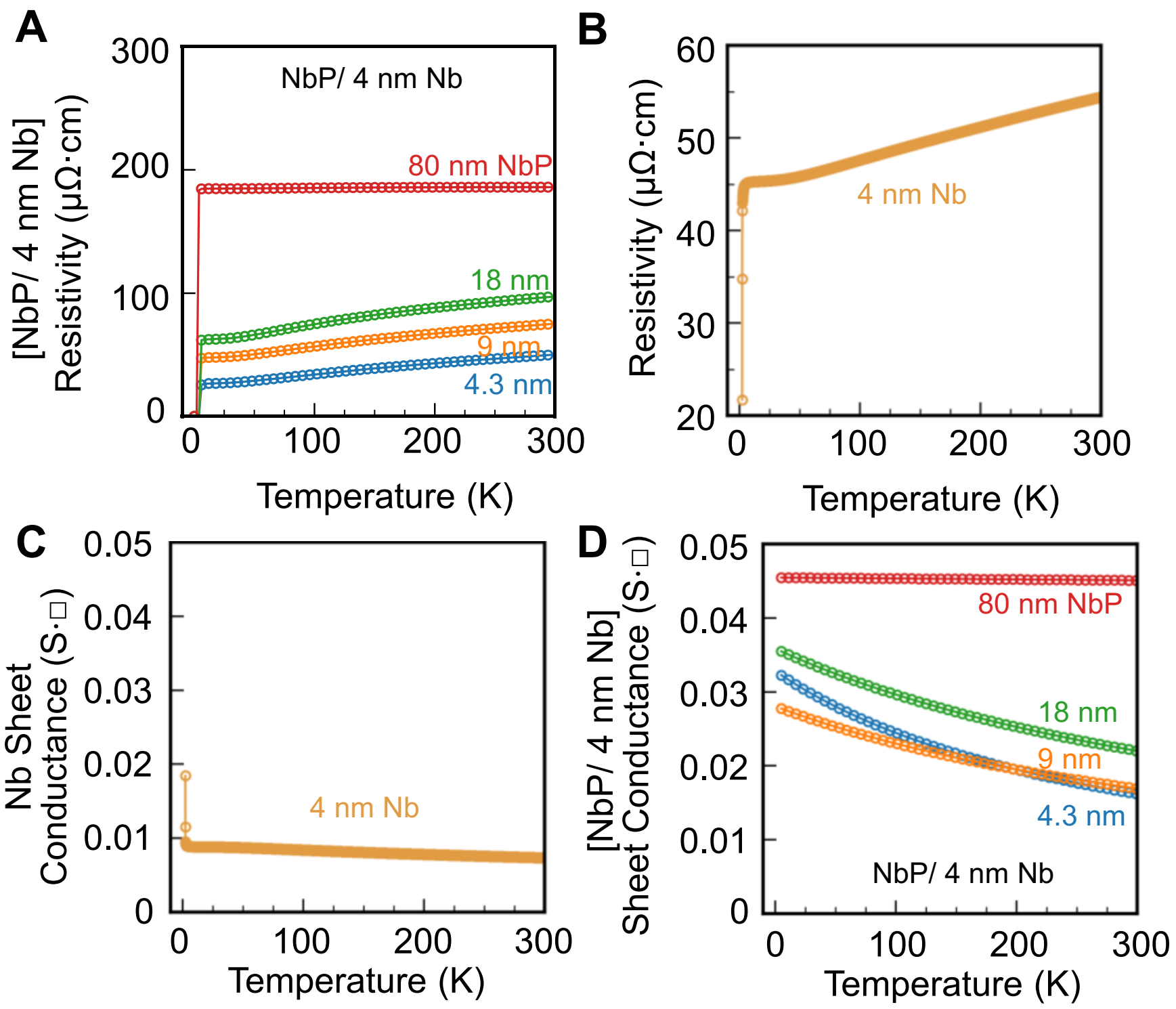

**Fig. S6. Temperature-dependent electrical measurement of Nb and NbP/Nb heterostructure**. Temperature dependent resistivity of **(A)** NbP/Nb films with varying NbP thickness (here 4.3, 9, 18, and 80 nm) on a 4 nm Nb seed, and **(B)** control 4 nm Nb film. Temperature-dependent sheet conductance of **(C)** control 4 nm Nb and **(D)** NbP/Nb films with varying NbP thickness (here 4.3, 9, 18, and 80 nm) on a 4 nm Nb seed. (Note, this is the same figure as Fig. 3C in the main manuscript, repeated here for convenience.) We note the resistivity in **A** is the total resistivity for the entire thickness of the sample (i.e., 8.3 nm to 84 nm), including the contribution of the 4 nm Nb seed layer. The control 4 nm Nb sample in **B,C** was prepared with the same deposition conditions as the 4 nm Nb seed layer under the NbP samples in **A,D**. The Nb seed layer is on the same sapphire substrate, capped by $SiN_x$ (see Materials and Methods, page 2 of this document).

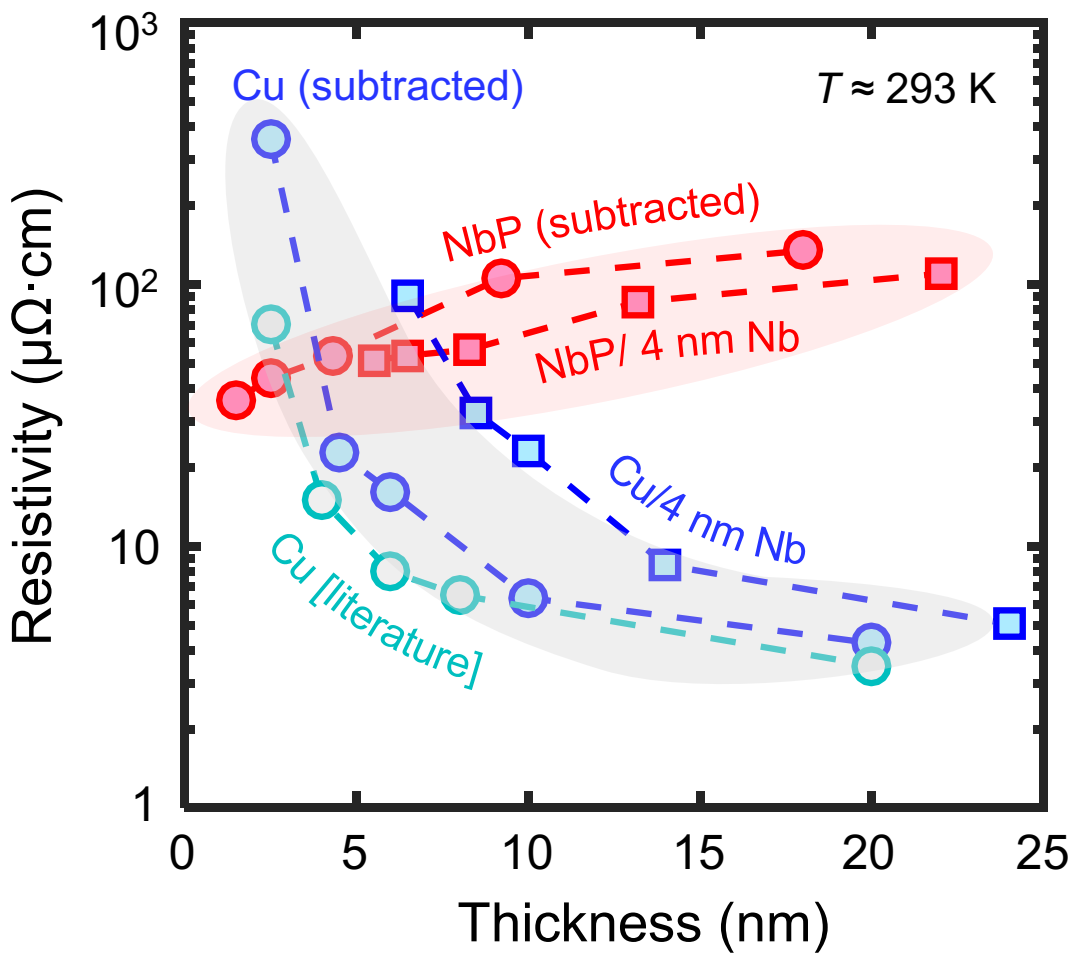

**Fig. S7. Electrical Resistivity of Cu/Nb and NbP/Nb stacks.** Room temperature resistivity vs. thickness of Cu/(4 nm Nb) heterostructures before (squares) and after (circles) subtracting the Nb seed layer conduction contribution. Unlike NbP/(4 nm Nb) heterostructures and NbP layers (after subtraction), the resistivity *increases* with decreasing total thickness for both Cu/(4 nm Nb) heterostructures and the Cu layers (using identical subtraction scheme). The resistivity of the Cu layer (after subtraction) as well as the resistivity vs. thickness trend are in agreement with the reported literature (*48*). We also note that in the Cu/Nb heterostructures, the bulk-like value nearly recovers the bulk resistivity of Cu (few μΩ·cm). Both types of films are capped *in situ* with the same $SiN_x$ layer (~3 nm) as described in Materials and Methods.

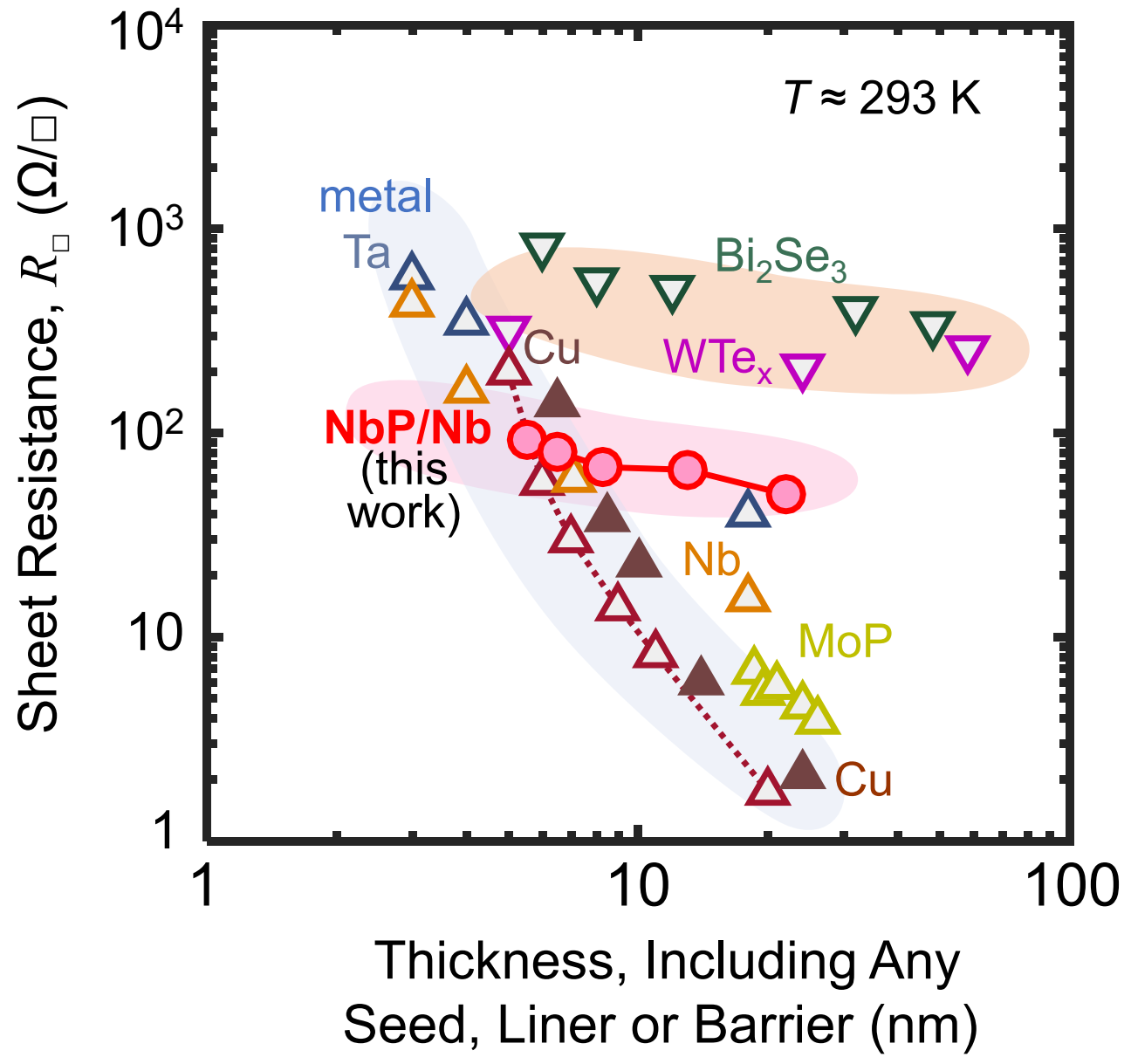

**Fig. S8.** Room temperature ($T \approx 293$ K) sheet resistance $R_\square$ vs. total thickness for various materials including our sputtered NbP semimetal (with 4 nm Nb seed), conventional metals like Cu (with liner and barrier, hollow triangles) (*48*), control Cu with Nb seed (from this work, filled triangles), Ta, Nb (from this work), other topological insulators (e.g., $Bi_2Se_3$) (*32*), topological semimetals (nanocrystalline $WTe_x$) (*23*, *26*), and a topological metal (MoP) (*45*) from the literature. Here, sheet resistance $R_\square = R_{meas}(W/L)$, where $R_{meas}$ is the measured resistance, $L$ and $W$ are the length and width of the samples, respectively. The sheet resistance of topological semimetals (NbP, $WTe_x$) and topological insulators ($Bi_2Se_3$) display a slowly increasing trend with decreasing thickness (shaded red and orange). In contrast, the sheet resistance of conventional metals increases much more strongly with decreasing thickness (shaded light blue trend), a bottleneck for future nanoelectronics. For a thickness decrease from ~20 nm down to ~5 nm, the sheet resistance of traditional metals increases by ~10× to 100×, whereas the sheet resistance of topological semimetals (and insulators) increases by only < 2×, demonstrating the unique potential of such materials in achieving low resistivity even at their ultra-scaled thicknesses.

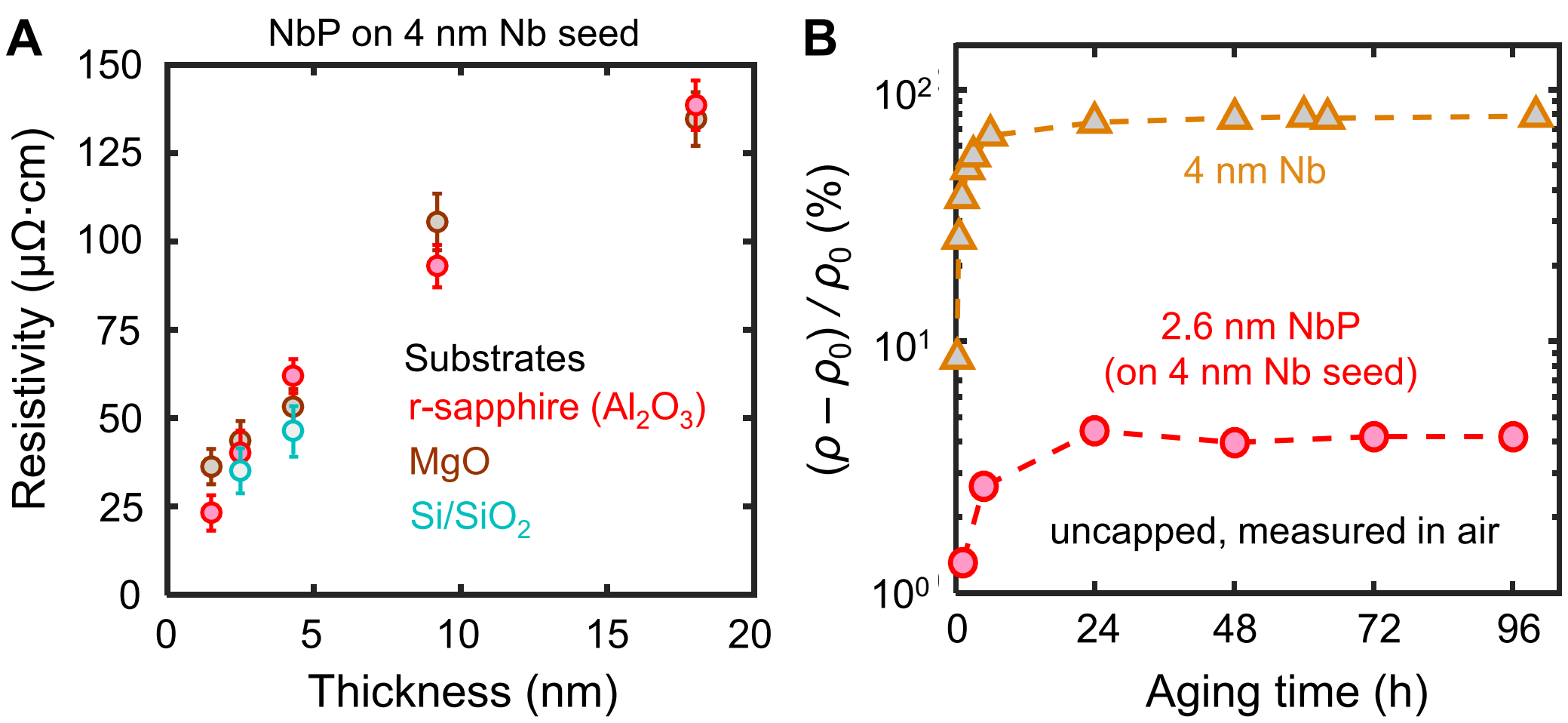

**Fig. S9. Electrical resistivity measurement of NbP. (A)** Resistivity versus thickness of NbP films on $Al_2O_3$ (sapphire), MgO and $SiO_2$/Si substrates. **(B)** Percentage change in the resistivity versus aging time for a ~2.6 nm thin NbP and a control 4 nm Nb metal films measured in air. $\rho_0$ is the resistivity measured immediately after deposition of the films, and $\rho$ is the resistivity measured after aging time steps. All the measurements (**A,B**) are taken at room temperature. NbP thin films were sputtered on 4 nm Nb seed layer. We subtracted the thickness and conductance contribution of the 4 nm Nb seed layer from the NbP/Nb stack.

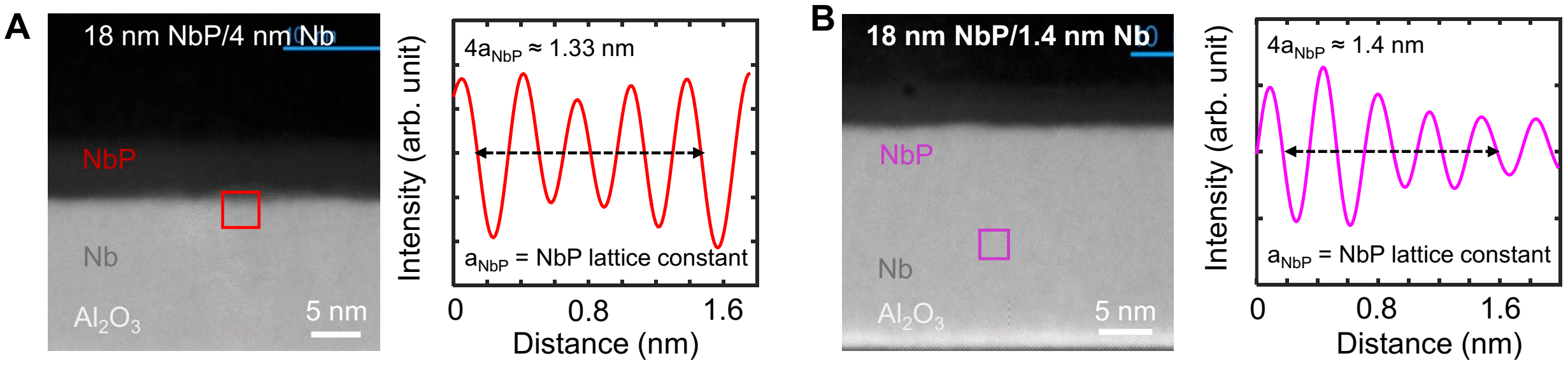

**Fig. S10. Microstructure details of NbP/Nb heterostructures.** Local nanocrystalline (short-range order) region of ~18 nm thin NbP film on **(A)** 4 nm Nb seed, showing NbP lattice constant ~3.33 Å, close to its nominal value of ~3.332 Å. **(B)** Similar image on 1.4 nm Nb seed layer, revealing NbP lattice constant ~3.5 Å, which indicates ~18 nm NbP is strained on the thinner Nb seed.

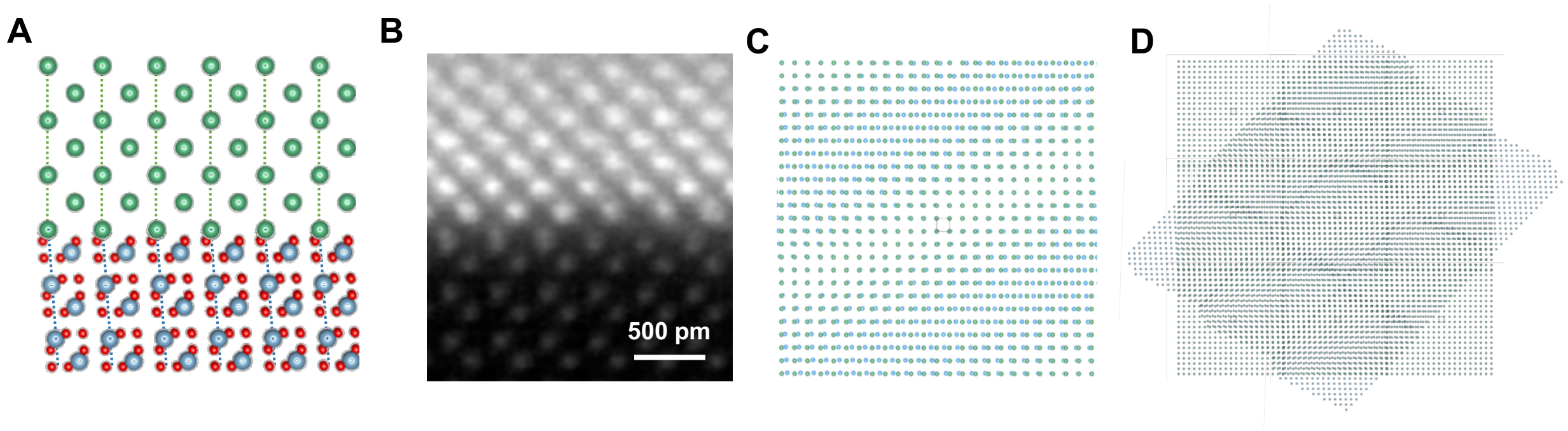

**Fig. S11. Strain and lattice constant of NbP on different Nb seed layers. (A)** Atomic projection of Nb [010] and $Al_2O_3$ [$\bar{2}11$]. **(B)** Atomic-resolution HAADF-STEM image of **A**. **(C)** Atomic projection of the Nb atoms in Nb (100) on Al atoms in $Al_2O_3$ (102). We assume that the distance of Nb-Nb (original distance = 3.32 Å) is the same as that of Al-Al (3.53 Å). **(D)** Wide projection view of (C) which shows the Moiré fringe. Nb in Nb (100) plane has a square lattice, while Al in α-$Al_2O_3$ (102) has a rhombus lattice tilted by $6^{o}$ compared to the square, which is clearly shown in the diffraction pattern using Fourier transform in Fig. 2E. Even if the atomic distance is the same, the coherency is periodically broken and misfit strain occurs in-plane, which is observed by Moiré pattern in projection of Nb (100) plane and α-$Al_2O_3$ (102) plane in fig. S8D. Thus, the interface between Nb (100) and α-$Al_2O_3$ (102) is semi-coherent interface and misfit dislocation should be introduced to release the strain energy as the thickness of Nb increases. The misfit dislocation is introduced at every 6 nm on the calculation by lattice mismatch and the Moiré distance due to lattice distortion is 3 nm. In our case, misfit dislocation was found at ~4 nm distance. Consequently, after the insertion of dislocation at ~1.5 nm thickness of Nb layer, the Nb lattice releases the compressive stress and returns to the original cubic structure with $a$ = 3.32 Å. In the case of the 1.4 nm Nb sample, misfit dislocation releasing the stress could not be observed within the Nb film, which means that the compressive stress due to lattice tensile ($a$ = 3.53 Å) remains in 1.4 nm Nb film.

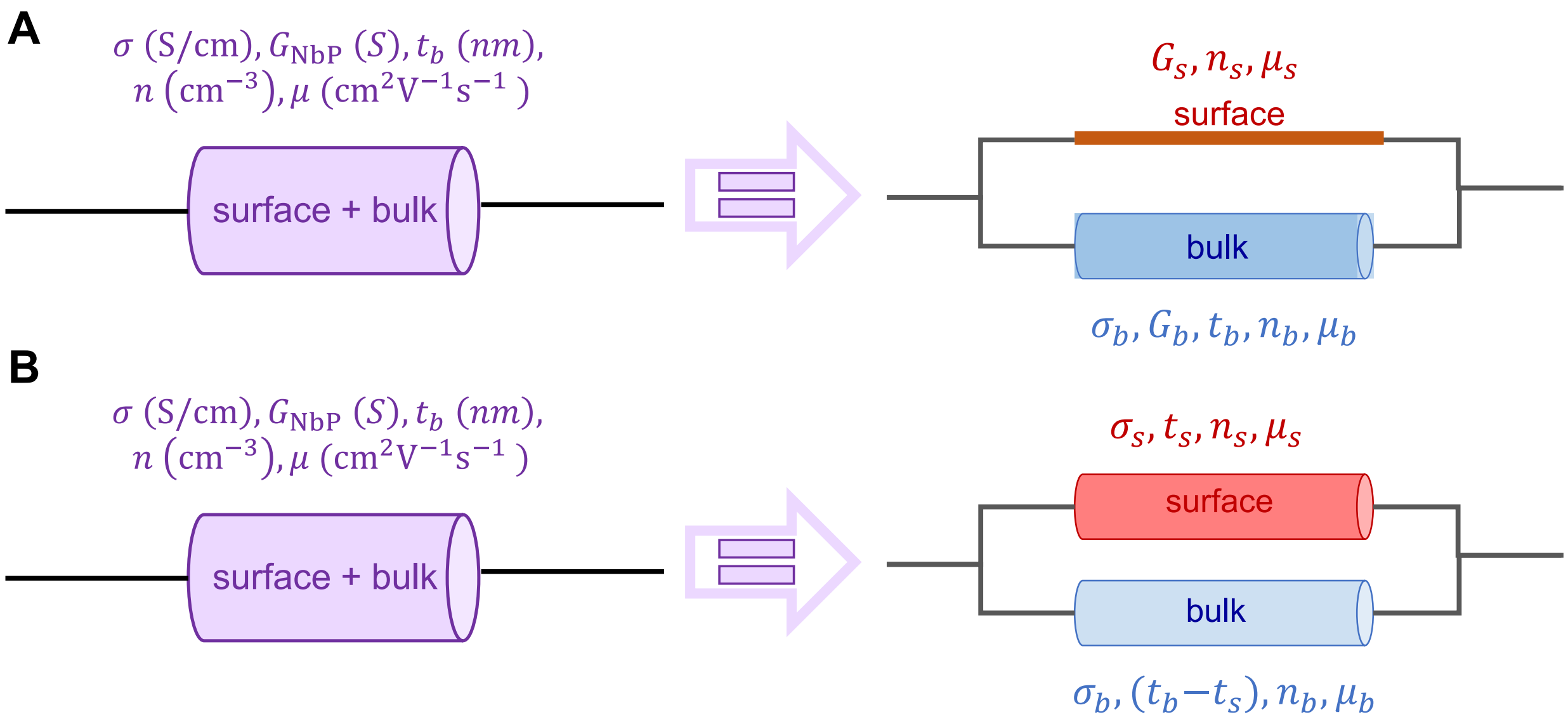

**Fig. S12**. **Surface- and bulk-channel conduction estimation for NbP**. The total sheet conductance, $G_{\mathrm{NbP}}$ of our NbP sample with a thickness $t_b$ can be modeled as $G_{\mathrm{NbP}} = G_b(t_b,T) + G_s(T)$, (**A**) Considering zero surface thickness, i.e., ideal two-dimensional (2D) surface with a sheet conductance $G_s$. (**B**) Considering a finite surface thickness $t_s$ = 5 Å with $G_{\mathrm{NbP}} = \sigma_b(T)(t_b - t_s) + G_s(T)$, where $G_s = \sigma_s t_s$. The 2D surface carrier density is $n_s$, the surface carrier mobility is $\mu_s$, $G_b$ is the bulk NbP conductance, $G_s$ is the NbP surface conductance, and $T$ is the temperature. The bulk conductivity is $\sigma_b$ (the inverse of resistivity, $1/\rho_b$) and the surface conductivity is $\sigma_s$.

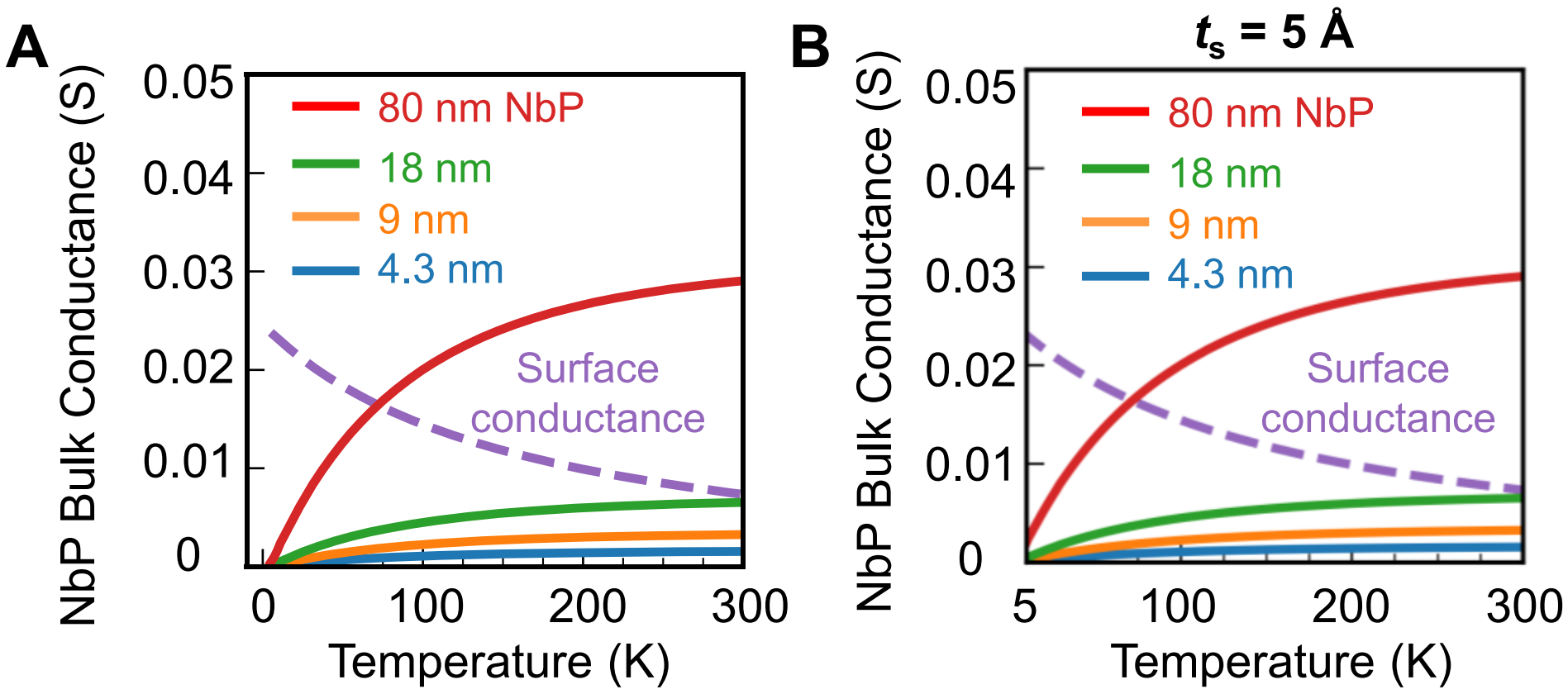

**Fig. S13**. **Bulk and surface conductance fits for NbP**. NbP bulk conductance and surface conductance for varying thicknesses of NbP films vs. temperature considering **(A)** an ideal 2D surface with zero surface thickness, and **(B)** a finite surface thickness $t_s = 5$ Å for NbP.

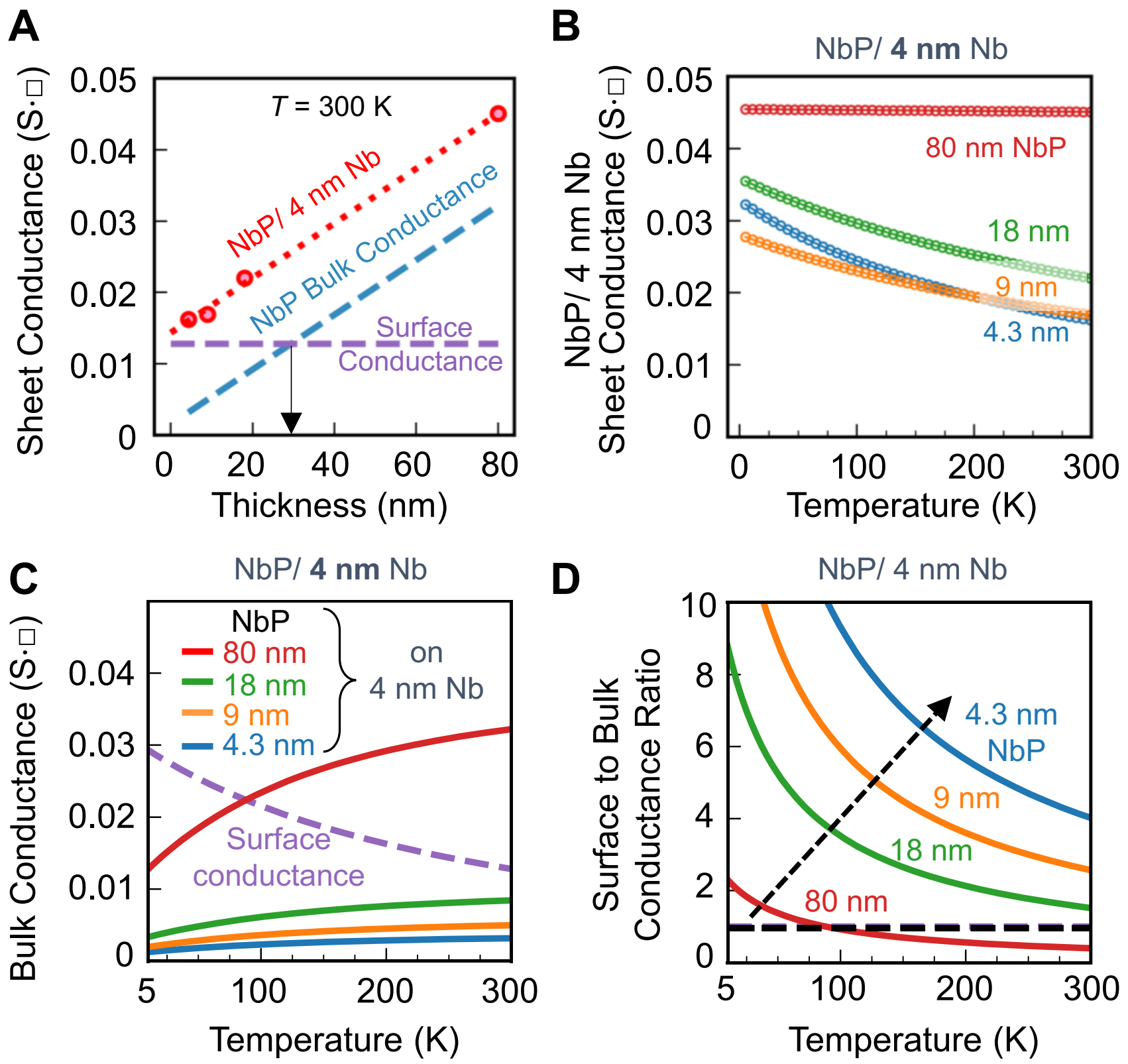

**Fig. S14**. **Temperature-dependent transport of NbP/Nb heterostructure**. **(A)** Room temperature sheet conductance of NbP/ 4 nm Nb heterostructures vs. NbP thickness. Red dotted line is a fit to the total sheet conductance of the NbP/ Nb heterostructures. NbP bulk conductance and 'effective surface conductance' (NbP surface conductance and 4 nm Nb seed conductance) are obtained through this fit (Materials and Methods section: Surface and Bulk Conductance of NbP/Nb and NbP Layer). The black arrow shows that the surface conductance dominates the total sheet conductance at room temperature for NbP/Nb films thinner than ~30 nm. **(B)** Temperature dependent sheet conductance of NbP / 4 nm Nb samples with varying NbP thicknesses (here 4.3, 9, 18, and 80 nm) on a 4 nm Nb seed. **(C)** Two-channel conductance fit (Materials and Methods: Surface and Bulk Conductance of NbP/Nb and NbP Layer) to the data in fig. S14B for various film thicknesses, indicating a metallic surface-channel (dashed line) and disorder dominated bulk channel conductance (solid lines). **(D)** Surface to bulk conductance ratio versus temperature for our NbP/Nb samples, showing that with decreasing film thicknesses, surface to bulk conductance ratio increases (indicated by the arrow). The region above the dashed line represents the surface conductance dominated area.

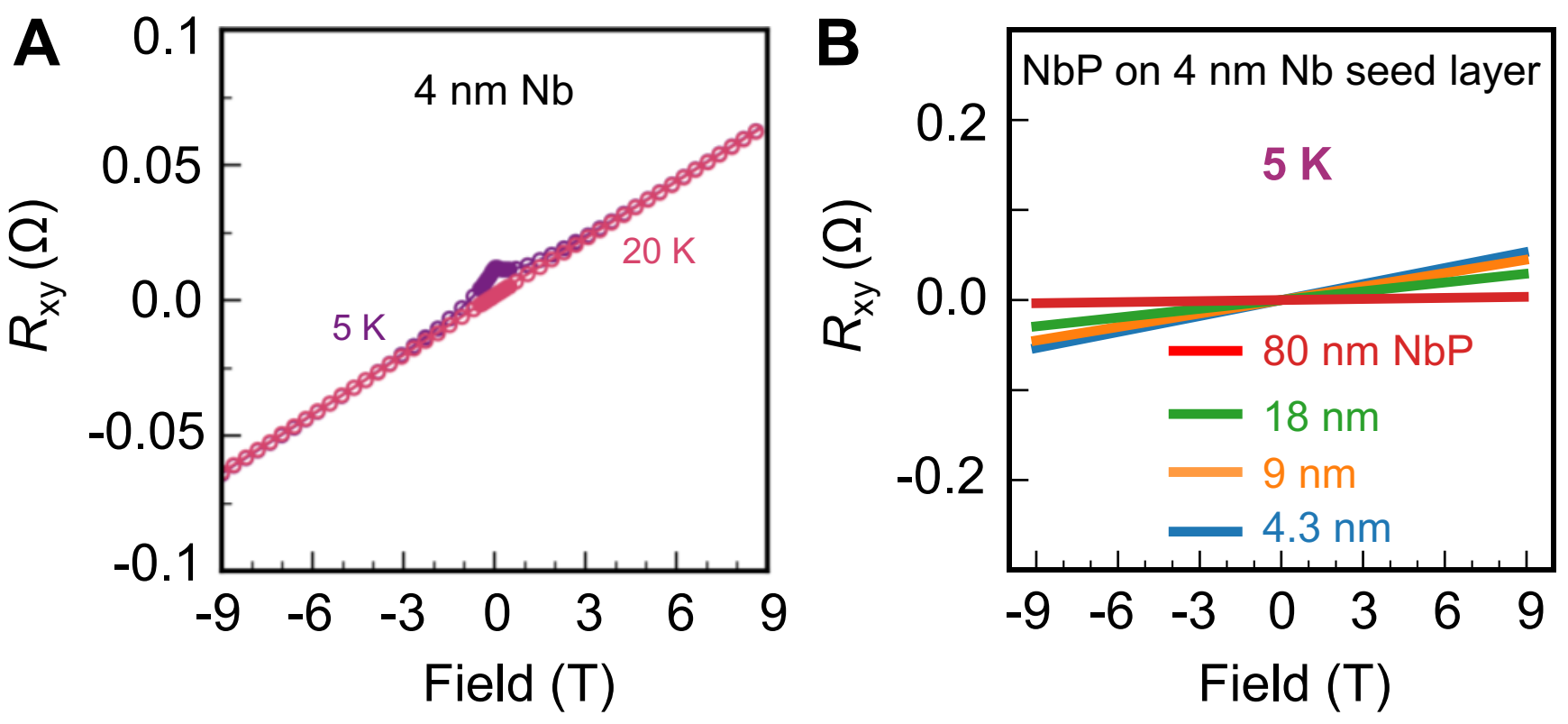

**Fig. S15. Magnetic field dependent Hall resistance measurements for Nb and NbP/Nb**. Hall resistance versus magnetic field of **(A)** control 4 nm Nb at 5 K and 20 K temperatures and **(B)** NbP/ 4 nm Nb samples for varying NbP thicknesses (80, 18, 9 and 4.3 nm) on a 4 nm Nb seed (at 5 K temperature). The control 4 nm Nb sample in **A** was prepared with the same deposition conditions as for the 4 nm Nb seed layer beneath the NbP samples in **B**.

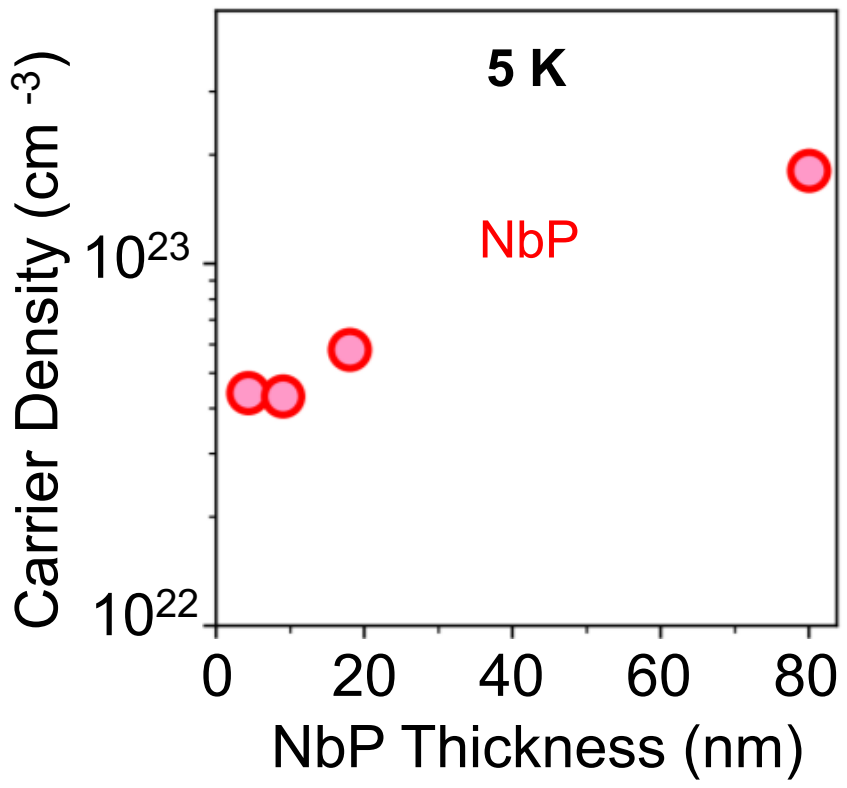

**Fig. S16. Total effective carrier density for NbP**. Estimated total carrier density (holes, per unit volume) (extracted from Fig. 4A) for various NbP film thicknesses. The corresponding sheet carrier density (in $cm^{-2}$) is shown in Fig. 4C. We note that total effective carrier density from Hall measurements in non-crystalline or disordered systems (as our non-crystalline NbP) could be overestimated (and the mobility underestimated) due to possible contribution from hopping-like transport (*46*). A similar observation has been reported in other systems such as organic semiconductors (*46*) and the topological insulator $Bi_2Se_3$, where the total carrier density in non-crystalline $Bi_2Se_3$ was estimated ~10× larger (*25*) compared to its crystalline counterpart (*32*).

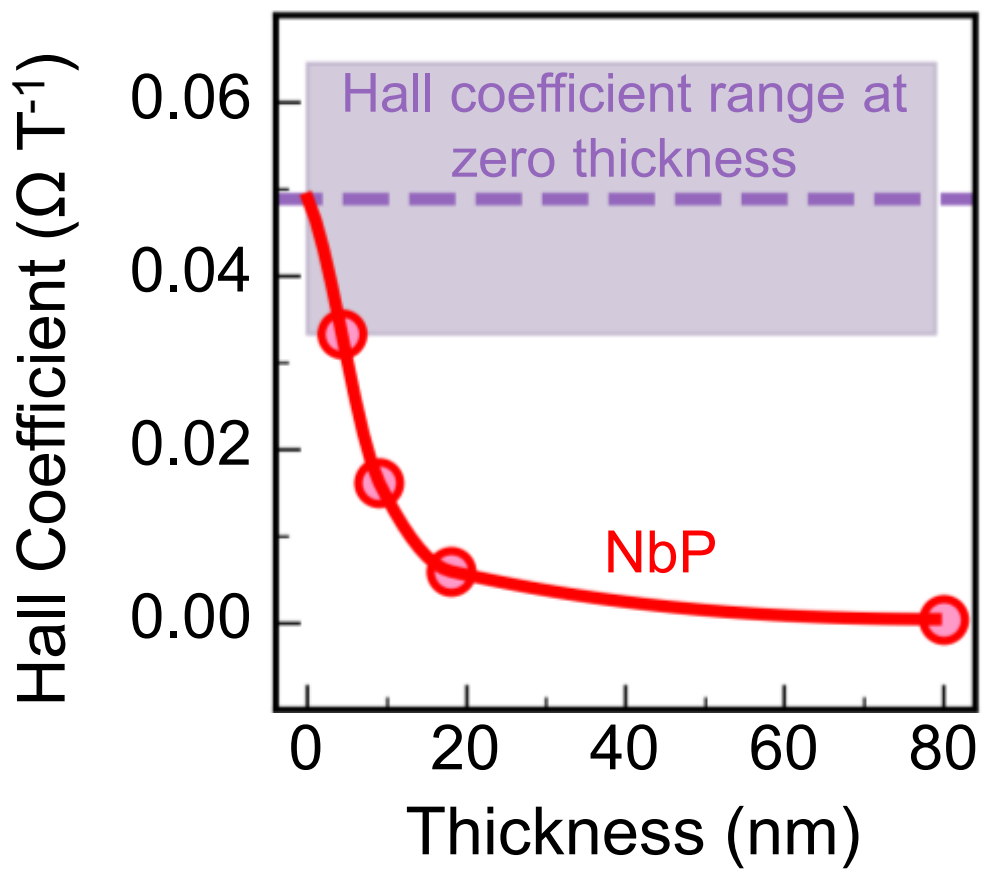

**Fig. S17. Hall coefficient measurements for NbP**. Measured Hall coefficient versus thickness of NbP films. Red line is a fit to the data extracted from measurements. Based on this fit, the purple dotted line represents the Hall coefficient when the NbP sample thickness → 0. As a conservative estimate for such a scenario, the Hall coefficient of the thinnest NbP sample (here, 4.3 nm) is defined as a lower bound (the bottom of the shaded purple region) for the Hall coefficient.

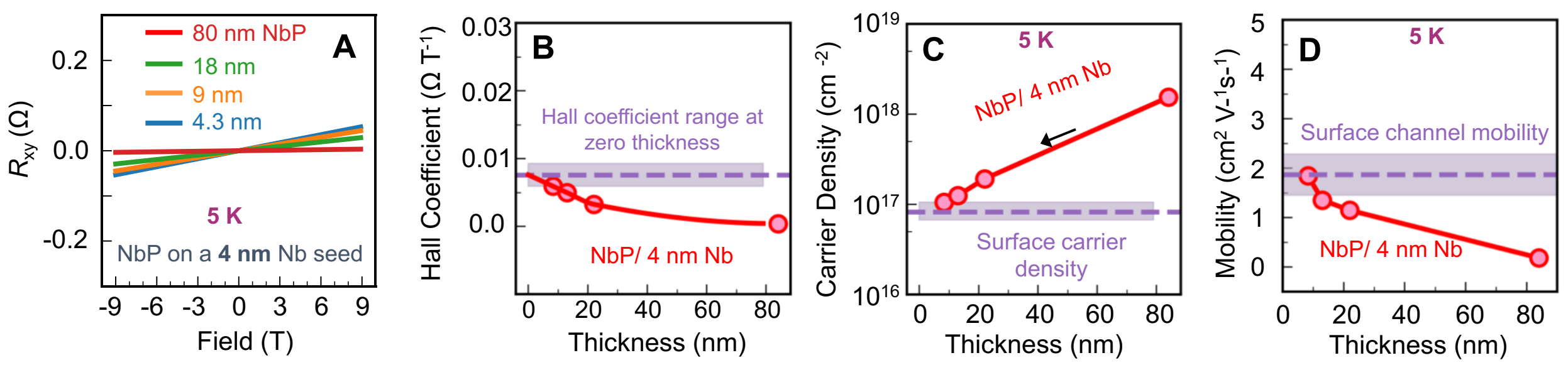

**Fig. S18**. **Magnetic field dependent transport of NbP/ 4 nm Nb stacks**. **(A)** Hall resistance versus magnetic field for NbP/ 4 nm Nb films at 5 K temperature. **(B)** Measured Hall coefficient versus thickness of NbP/ 4 nm Nb film stacks. Red line is a fit to the data extracted from measurements. Based on this fit, the purple dotted line represents the Hall coefficient when the NbP/4 nm Nb stack total thickness tends to 0 (similar approach as in **fig. S17**). **(C)** Two-dimensional (sheet) carrier density (extracted from fig. S17A,B) showing a decrease in the carrier density with decreasing NbP/4 nm Nb total film thicknesses. The shaded purple region represents the sheet carrier density in the limit of zero NbP/ 4 nm Nb total film thickness. **(D)** Mobility of the NbP/ 4 nm Nb samples, showing an increasing trend with decreasing total sample thicknesses. The shaded region represents the range of the surface channel mobility, estimated from the surface carrier density. As the films get thinner, the total mobility approaches the surface channel mobility. Thus, the inclusion of the 4 nm Nb seed layer conductivity contribution to the NbP films does not alter the trends discerned in Fig. 4. We note that carrier density estimated from Hall measurements in non-crystalline or disordered systems could be overestimated (and the mobility underestimated) due to possible contribution from hopping-like transport (*46*).

| Material | Power (W) | Pressure (mTorr) | Gas flow (sccm) | Temperature |
|---|---|---|---|---|
| **Nb (seed layer)** | 30 (DC) | 3 | Ar: 20 | 400 °C |
| **NbP** | 15 (DC) | 3 | Ar: 20 | 400 °C |
| **$SiN_x$ (capping layer)** | 100 (RF) | 4 | Ar: 30 | Room temperature |

**Table S1. Materials deposition parameters.** Sputtering parameters for various materials used in this work. DC: direct current, RF: radio frequency.